%% file: island_complexity.tex
\documentclass[onecolumn,amsmath,amssymb,nofootinbib,12pt]{article}
\pdfoutput=1 

\usepackage{jheppub} 

\usepackage{graphicx,subcaption}
\graphicspath{ {figures/} }
\usepackage{amsfonts}
\usepackage{xparse}
 \usepackage{amsmath}
 \usepackage{amssymb}
 \usepackage{mathtools}
 \usepackage{tensor}
\usepackage{epsfig}
\usepackage{pdfpages}
\usepackage{bbm}
\usepackage{graphicx,epstopdf}
\usepackage[numbers]{natbib}
\usepackage[makeroom]{cancel}
\usepackage{hyperref}
\hypersetup{pdftex,colorlinks=true,allcolors=blue}
\usepackage{array}
\usepackage[export]{adjustbox}
\usepackage[normalem]{ulem}
\usepackage{tcolorbox}

\newcommand{\eg}{{\it e.g.,}\ }
\newcommand{\ie}{{\it i.e.,}\ }
\newcommand{\mt}[1]{\textrm{\tiny #1}}
\newcommand{\reef}[1]{(\ref{#1})}

\newcommand{\beq}{\begin{equation}}
\newcommand{\eeq}{\end{equation}}
\newcommand{\beqa}{\begin{eqnarray}}
\newcommand{\eeqa}{\end{eqnarray}}

\newcommand{\beqs}{\begin{equation}\begin{aligned}}
\newcommand{\eeqs}{\end{aligned}\end{equation}}

\renewcommand{\(}{\left(}
\renewcommand{\)}{\right)}
\renewcommand{\[}{\left[}
\renewcommand{\]}{\right]}

\newcommand{\veps}{\varepsilon}
\newcommand{\ads}[1]{{\mt{AdS}_{\tiny #1}}}
\newcommand{\mC}{\mathcal{C}}
\newcommand{\Gbulk}{G_\mathrm{bulk}}
\newcommand{\Gbrane}{G_\mathrm{brane}}

\newcommand{\Geff}{G_\mathrm{eff}}
\newcommand{\hGeff}{\widehat{G}_\mathrm{eff}}

\newcommand{\mR}{\mathcal{R}}
\newcommand{\RB}{R_\mathcal{B}}
\newcommand{\mK}{\mathcal{K}}
\newcommand{\mL}{\mathcal{L}}
\newcommand{\mB}{\mathcal{B}}
\newcommand{\wmB}{\widetilde\mB}
\newcommand{\mS}{\mathbf{S}}
\newcommand{\mA}{\mathbf{R}}

\newcommand{\tn}{\tilde{n}}
\newcommand{\tg}{\tilde{g}}
\newcommand{\tR}{\tilde{R}}
\newcommand{\tK}{\tilde{K}}

\newcommand{\tL}{\tilde{L}}
\newcommand{\tdh}{\tilde{h}}
\newcommand{\zB}{z_{\mt{B}}}
\def\ov#1#2{\overset{(#1)}{#2}}

\newcommand{\bdyReg}{\mathbf{R}}

\newcommand{\RTbrn}{\sigma_{\bdyReg}}

\newcommand{\RT}{\Sigma_{\bdyReg}}

\newcommand{\islands}{\mathrm{islands}}
\newcommand{\area}{A}
\newcommand{\bulk}{\text{bulk}}
\newcommand{\brane}{\mathrm{brane}}
\title{Quantum Extremal Islands Made Easy, Part \!I\!\,I\!\,I:\\
Complexity on the Brane}

\author[a,b]{Juan Hernandez,}
\author[a]{Robert C. Myers,}
\author[a,b]{and Shan-Ming Ruan}

\affiliation[a]{Perimeter Institute for Theoretical Physics, Waterloo, ON N2L 2Y5, Canada}
\affiliation[b]{Dept.~of Physics $\&$ Astronomy, University of Waterloo, Waterloo, ON N2L 3G1, Canada}

\date{\today}

\abstract{We examine holographic complexity in the doubly holographic model introduced in \cite{Chen:2020uac,Chen:2020hmv} to study quantum extremal islands. We focus on the holographic complexity=volume (CV) proposal for boundary subregions in the island phase. Exploiting the Fefferman-Graham expansion of the metric and other geometric quantities near the brane, we derive the leading contributions to the complexity and interpret these in terms of the generalized volume of the island derived from the induced higher-curvature gravity action on the brane. Motivated by these results, we propose a generalization of the CV proposal for higher curvature theories of gravity. Further, we provide two consistency checks of our proposal by studying Gauss-Bonnet gravity and $f(\mR)$ gravity in the bulk.}

\begin{document}

\maketitle

\section{Introduction}\label{sec:intro}
\input{sections/introduction}

\section{Holographic complexity on the island}\label{sec:bulksubregion}
\input{sections/SubVolume}

\section{Higher curvature gravity in the bulk}\label{sec:GB}
\input{sections/Gauss}

\section{Discussion and future directions}\label{sec:dis}
\input{sections/Discussion}

\section*{Acknowledgments}
We would like to thank Alex Belin, Shira Chapman, Dominik Neuenfeld and Antony Speranza for useful comments and discussions. Research at Perimeter Institute is supported in part by the Government of Canada through the Department of Innovation, Science and Economic Development Canada and by the Province of Ontario through the Ministry of Colleges and Universities. RCM is supported in part by a Discovery Grant from the Natural Sciences and Engineering Research Council of Canada, and by the BMO Financial Group.  RCM also received funding from the Simons Foundation through the ``It from Qubit'' collaboration. JH is also supported in part by the Natural Sciences and Engineering Research Council of Canada through a Postgraduate Doctoral Scholarship. 

\appendix 

\section{Lower dimensions} \label{app:lower}
\input{sections/lowerD23}


\bibliographystyle{JHEP}
\bibliography{references}

\end{document}

%% file: sections/introduction.tex

In the last few years, an influx of concepts from quantum information theory have led to exciting new insights about quantum gravity, especially within the framework of gauge/gravity duality~\cite{Maldacena:1997re}. One of these concepts that has been a topic of much research is the quantum circuit complexity~\cite{Aaronson:2016vto}, which quantifies how difficult it is to prepare a target state from a simple reference state, given a particular set of elementary gates. Among the various conjectured holographic duals to circuit complexity, the two most extensively studied are the complexity=volume (CV)~\cite{Susskind:2014rva,Stanford:2014jda} and the complexity=action (CA)~\cite{Brown:2015bva,Brown:2015lvg} proposals. The CV conjecture states that the complexity of the state in the boundary theory defined on a time slice $\mS$ is dual to the volume of the maximal codimension-one bulk surface anchored to $\mS$ on the asymptotic boundary,
\begin{equation}\label{eq:CVpure}
{\cal C}_{\mt{V}}(\mS)= \max_{\partial\mB=\mS} \[\frac{V(\cal B)}{G_{\mt{N}} \,\ell} \]\,,
\end{equation}
where $G_{\mt{N}}$ is the Newton's constant of bulk gravity theory and $\ell$ is some undetermined length scale. Various aspects of the CV proposal have been studied on the gravitational side of the duality, \eg see  \cite{Carmi:2016wjl,Couch:2016exn,Carmi:2017jqz,Swingle:2017zcd,Chapman:2018dem,Chapman:2018lsv,Fu:2018kcp,Flory:2018akz,Bernamonti:2020bcf,Chen:2020nlj,Sarkar:2020yjs,Couch:2018phr,Balushi:2020wjt,Cai:2020wpc,Jian:2020qpp}. The above conjecture assumes that the state in question is a pure state defined on a global time slice, \ie the time slice $\mS$ spans the entire asymptotic boundary. 

Motivated by entanglement wedge reconstruction~\cite{Czech:2012bh,Wall:2012uf,Headrick:2014cta,Jafferis:2015del,Dong:2016eik,Cotler:2017erl}, the CV proposal was extended to mixed states produced by reducing a global pure state down to a subregion of the boundary~\cite{Alishahiha:2015rta,Carmi:2016wjl}. The subregion-CV conjecture proposes that the complexity of the quantum state defined on a boundary subregion $\mA$ is given by the volume of a maximal codimension-one bulk surface extending from  $\mA$ on the asymptotic boundary to the corresponding Ryu-Takayanagi (RT) surface $\RT$ in the bulk,
\begin{equation}\label{eq:CVsubregion}
{\cal C}^{\rm sub}_{\mt{V}} ( \mA ) = \max_{ \partial{\mB}=\mA\,\cup\,\RT} \[\frac{V(\mB)}{G_{\mt{N}} \,\ell} \]\,.
\end{equation}
For example, see  \cite{Alishahiha:2015rta,Abt:2017pmf,Bakhshaei:2017qud,Abt:2018ywl,Caceres:2018luq,Chapman:2018bqj,Agon:2018zso,Chen:2018mcc,Bhattacharya:2018oeq,Cooper:2018cmb,Bhattacharya:2019zkb,Karar:2019bwy,Auzzi:2019fnp,Lezgi:2019fqu,Auzzi:2019mah,Ling:2019ien,Chakrabortty:2020ptb,Ben-Ami:2016qex,Caceres:2018blh,Caceres:2019pgf,Auzzi:2019vyh,Braccia:2019xxi,Sato:2019kik} for more recent explorations on the subregion-CV proposal.

Recently, information theoretic ideas have also produced exciting new insights for the resolution of the black hole information paradox~\cite{Hawking:1974sw,Hawking:1976ra,Page:1993wv}. The latter can be quantified by examining the von Neumann entropy of the Hawking radiation \cite{Page:1993wv,Page:2013dx,Harlow:2014yka}. Hawking's original analysis indicated that this entropy increases throughout the evaporation of a black hole since one is simply accumulating more and more thermal radiation. However, Page argued that the entropy of the radiation must be bounded by the black hole entropy for a unitary evolution, so the entropy must in fact decrease over the second half of the evaporation process and reach zero in the final state where the black hole has disappeared. The {\it Page curve} is then a plot of the entropy of the Hawking radiation as a function of time which exhibits this qualitative behaviour \cite{Page:1993wv,Page:2013dx}.

Recent progress~\cite{Almheiri:2020cfm,Almheiri:2019psf,Penington:2019npb, Almheiri:2019hni}\footnote{These calculations were subsequently applied in a broad variety of situations, \eg see 
  \cite{Chen:2019uhq,Rozali:2019day,Chen:2020uac,Chen:2020hmv,Gautason:2020tmk,Sully:2020pza,Chen:2019iro,Anegawa:2020ezn,Balasubramanian:2020hfs,Hartman:2020swn,Hollowood:2020cou,Alishahiha:2020qza,Almheiri:2019psy,Hashimoto:2020cas,Geng:2020qvw,Bak:2020enw,Li:2020ceg,Chandrasekaran:2020qtn, Hollowood:2020kvk,Almheiri:2019qdq,Bousso:2019ykv,Penington:2019kki,Akers:2019nfi,Chen:2020wiq,Kim:2020cds,Verlinde:2020upt,Liu:2020gnp,Bousso:2020kmy, Balasubramanian:2020coy, Chen:2020jvn, Stanford:2020wkf,Marolf:2020xie,Hartman:2020khs,Giddings:2020yes,Chen:2020tes,VanRaamsdonk:2020tlr, Sybesma:2020fxg, Balasubramanian:2020xqf,Bhattacharya:2020uun,Ling:2020laa}.}   into understanding  the Page curve builds on insights coming from holographic entanglement entropy \cite{Ryu:2006ef,Ryu:2006bv,Hubeny:2007xt,Lewkowycz:2013nqa,Dong:2016hjy,Rangamani:2016dms} and its extension to include quantum contributions \cite{Faulkner:2013ana,Engelhardt:2014gca}. For simplicity, one assumes that the Hawking radiation is absorbed by a non-gravitational reservoir (the {\it bath}), which is coupled to the asymptotic boundary of the gravitational region containing the black hole. One finds   the entropy of  the Hawking radiation in a bath subregion $\mA$ is given by the \emph{island rule} \cite{Almheiri:2019hni,Almheiri:2020cfm} 
\beq\label{eq:islandformula}
 S_\mt{EE}(\bdyReg) = \text{min} \left\{\underset{\islands}{\text{ext}}  \left(S_\mt{QFT}( \bdyReg \cup \islands ) +\frac{A\!\left( \partial( \islands )\right)}{4G_{\mt N}}   \right)  \right\}\,.
\eeq
That is, $S_\mt{EE}(\bdyReg)$ is not just given by the entropy of the quantum fields in the bath region, but rather one also considers $\mA$ together with subregions (\ie {\it islands}) in the gravitating region to minimize the entanglement entropy of the combined subregion. Further, the Bekenstein-Hawking entropy appears as an additional gravitational contribution at the boundary of the islands.  

Initially, for an evaporating black hole, eq.~\reef{eq:islandformula} is minimized without any islands and the calculation matches Hawking's evaluation of the entropy. However, at late times, a new saddle point involving a nontrivial island dominates because the Hawking radiation shares a large amount of entanglement with the quantum fields behind the horizon. In this Page phase of the time evolution, the entropy is controlled by the black hole entropy, which appears in the second term in eq.~\eqref{eq:islandformula}, and in this way, the island rule yields the expected unitary Page curve.

The island rule has a simple interpretation within certain ``doubly-holographic'' models in \cite{Almheiri:2019hni,Rozali:2019day,Chen:2020uac,Chen:2020hmv}. Of course, the physics can be described with the usual bulk and boundary perspectives of a holographic system. In this case, the \emph{boundary perspective} consists of a $d$-dimensional CFT coupled to a codimension-one conformal defect, and the \emph{bulk perspective} then becomes ($d$+1)-dimensional gravity on an asymptotically AdS spacetime containing a codimension-one brane, which is anchored at the conformal defect on the asymptotic boundary. This brane back reacts on the bulk spacetime and in an appropriate parameter regime, a third perspective emerges through the Randall-Sundrum mechanism \cite{Randall:1999ee,Randall:1999vf,Karch:2000ct}. In this \emph{brane perspective}, the brane supports a theory of $d$-dimensional gravity coupled to (two copies of) the holographic CFT,  and is connected to the CFT on the asymptotic boundary (which becomes the bath) at the position of the defect. We refer the interested reader to \cite{Chen:2020uac,Chen:2020hmv} for further details on these three perspectives.

\begin{figure}[t]
	\centering
		\includegraphics[width=5.8in]{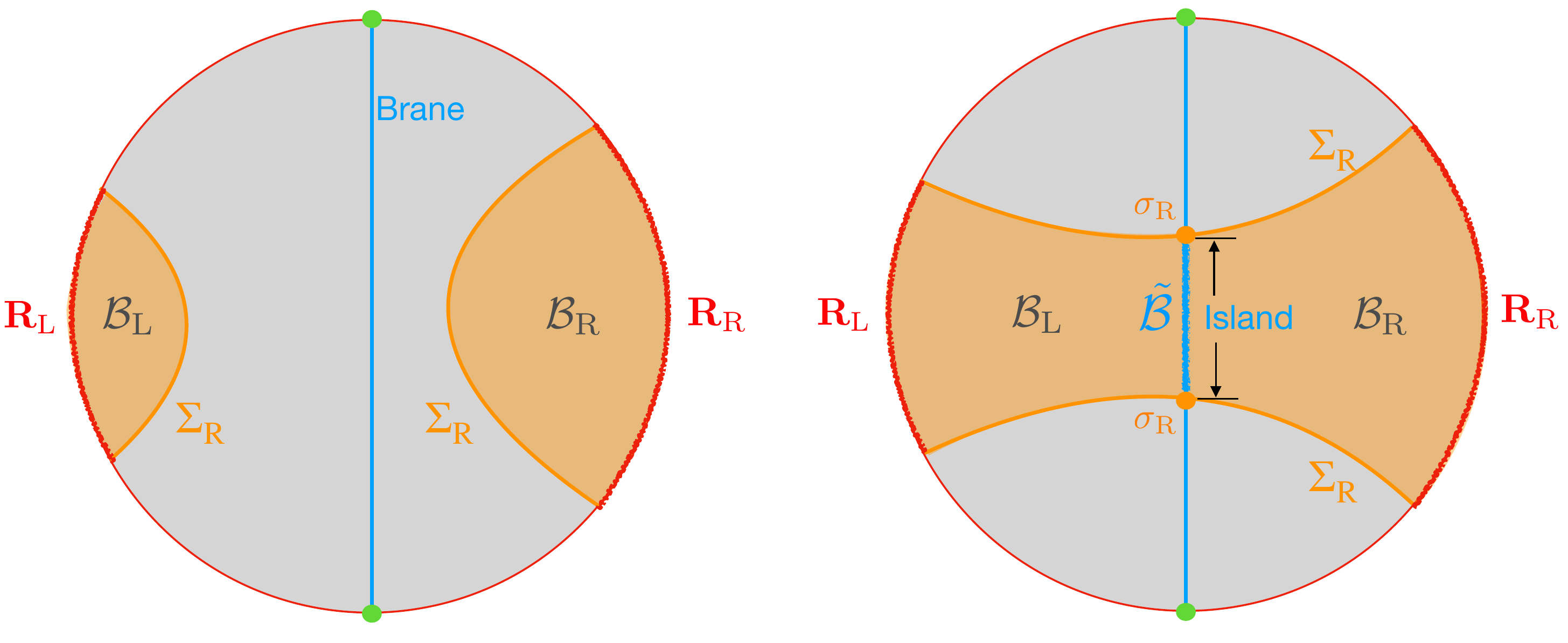}
		\caption{The choice of RT surfaces for the boundary subregion $\mA=\mA_{\mt{L}}\cup\mA_{\mt{R}}$ on a constant time slice in the presence of the brane (coloured green), showing the island and no-island phases in the right and left panels, respectively. The  complexity ${\cal C}^{\rm sub}_{\mt{V}}(\mA)$ in eqs.~\eqref{eq:CVsubregion} and \eqref{eq:CVsubregion2} is determined by the extremal surface $\mB=\mB_{L}\cup \mB_R$. In the island phase, the intersection of this surface with the brane defines the `island' $\wmB=\mB\,\cap\,$brane.}
\label{fig:simple}		
\end{figure}

A key advantage of this framework is that entanglement entropies in eq.~\reef{eq:islandformula} are calculated purely geometrically from the bulk perspective, using the usual rules of holographic entanglement entropy \cite{Ryu:2006ef,Ryu:2006bv,Hubeny:2007xt,Lewkowycz:2013nqa,Dong:2016hjy,Rangamani:2016dms}. In particular, the entanglement entropy for a bath or boundary region $\bdyReg$ becomes
\begin{equation}\label{eq:island2}
 S_\text{EE}(\bdyReg)= \text{min} \left\{\underset{\RT}{\text{ext}}  \left( \frac{\area(\RT)}{4 G_\bulk}
  + \frac{\area(\RTbrn)}{4 G_\brane}  \right)  \right\}\,,
\end{equation}
where $\RT$ is the usual bulk RT surface, while $\RTbrn=\RT\cap \text{brane}$ is the intersection of the RT surface with the brane. The second term in eq.~\eqref{eq:island2} is the Bekenstein-Hawking area contribution that is included when an intrinsic gravitational action (\ie a DGP term \cite{Dvali:2000hr}) is included in the brane action~\cite{Chen:2020uac,Chen:2020hmv}. From the brane perspective then, islands simply arise when the minimal RT surfaces in the bulk extend across the brane, as illustrated in the right panel of figure \ref{fig:simple}. Further, the transition between the island and no-island phases (\eg between the Page and Hawking phases of an evaporating black hole) corresponds to a conventional transition found in holographic entanglement entropy between different classes of RT surfaces, \eg \cite{Headrick:2010zt,Faulkner:2013yia,Hartman:2013mia,Belin:2017nze}. Let us add that carefully examining eq.~\eqref{eq:island2} near the brane shows that the gravitational contribution in the island rule \eqref{eq:islandformula} expands to the Wald-Dong entropy \cite{Wald:1993nt,Iyer:1994ys,Jacobson:1993vj,Dong:2013qoa} for the higher-curvature  gravitational action induced on the brane~\cite{Chen:2020uac}. 

In this paper, we extend the examination of the model constructed in~\cite{Chen:2020uac,Chen:2020hmv} to consider holographic complexity, and in particular, the subregion-CV proposal \reef{eq:CVsubregion}. In particular, we focus on the island phase (\ie the right panel of figure \ref{fig:simple}) in which case the extremal bulk surface $\mB$ also crosses the brane. Following an analysis similar to that of~\cite{Chen:2020uac} for the holographic entanglement entropy, we employ the FG expansion of the subregion-CV in the vicinity of the brane to recast it as an integral of geometric quantities over the island, \ie $\widetilde\mB=\mB\,\cap\,$brane. Then  to leading order, eq.~\reef{eq:CVsubregion} yields
\begin{equation}\label{eq:CVsubregion2}
{\cal C}^{\rm sub}_{\mt{V}} ( \mA ) \simeq {\rm max} \[\frac{d-2}{d-1}\,\frac{V(\widetilde\mB)}{G_{\mt{eff}} \,\ell} +\cdots \]\,,
\end{equation}
where $G_\mt{eff}$ is the induced Newton's constant for the gravitational theory on the brane.\footnote{Note that here we are ignoring the UV divergent contributions coming from where $\mB$ approaches the asymptotic boundary. This result also assumes there is no DGP term on the brane.} Setting aside the dimension-dependent prefactor, the geometric integral over $\widetilde\mB$ is naturally interpreted as the holographic complexity of the island region.

However, beyond the volume term, the ellipsis in eq.~\reef{eq:CVsubregion2} also includes higher curvature corrections. By examining these contributions, we are lead to a generalized CV formula derived from the induced higher-curvature gravity action on the brane. That is, we propose to generalize the complexity=volume conjecture for an arbitrary $(d+1)$-dimensional higher-curvature gravity theory in the bulk as   
\begin{equation}\label{eq:ourporposal}
\mC_\mt{V}(\mA)=\max_{\partial\mB=\mA} \[  \frac{W_{\rm gen}(\mathcal{B}) + W_K(\mathcal{B})}{G_\mt{N}\,\ell}  \] \,.  \qquad (d>2)
\end{equation}
where $ W_{\rm gen}$ is called the generalized volume because this expression reduces to the volume term $V(\mB)$ for Einstein gravity, and $W_K$ introduces extra corrections involving the extrinsic curvature $\mK_{\mu\nu}$ of the hypersurface $\mB$. Explicitly, our analysis determines these two contributions as 
\begin{equation}\label{eq:details}
\begin{split}
W_{\rm gen}(\mathcal{B}) =&\frac{2}{(d-1)(d-2)}\int_{\mB} d^{d}\sigma\, \sqrt{\det h} \left(1+(d-3)\, \frac{\partial \mathbf{L}_{\rm bulk}}{\partial \mR_{\mu\nu\rho\sigma}} n_{\mu} h_{\nu\rho} n_{\sigma}\right) \,,\\
W_K(\mathcal{B})  =&\frac{4(d-3)}{(d-1)^2(d-2)} \int_{\mB} d^{d}\sigma\,\sqrt{\det h}\  \frac{\partial^2  \mathbf{L}_{\rm bulk}}{\partial \mR_{\mu_1\nu_1\rho_1\sigma_1}\partial \mR_{\mu_2\nu_2\rho_2\sigma_2}}  \\
& \times\ \left[ \mK_{\nu_1\sigma_1}\(  h_{\mu_1\rho_1} + (d-2) n_{\mu_1} n_{\rho_1}\)\, \mK_{\nu_2\sigma_2} \(h_{\mu_2\rho_2} + (d-2) n_{\mu_2} n_{\rho_2}\) \]+\cdots\,.
\end{split}
\end{equation}
For these expressions, we have rescaled the gravitational Lagrangian so that the gravitational action carries an overall factor: $I_\mt{grav}=\frac{1}{16 \pi G_\mt{N}}\int d^{d+1}x \sqrt{-g}\, \mathbf{L}_{\rm bulk}$. Further, $\mB$ denotes a spacelike codimension-one bulk hypersurface with unit normal $n^\mu$, induced metric $h_{\mu\rho}$, and extrinsic curvature $\mK_{\mu\nu}$. The generalized subregion-CV functional is maximized subject to the constraint that the codimension-one hypersurface $\mB$ is anchored at the boundary subregion $\mA$ and the corresponding RT surface $\RT$, \ie $\partial {\cal B} = \mA\cup \RT$. 

Our proposal to the generalized CV contains two contributions, in a similar spirit to the Wald-Dong entropy \cite{Wald:1993nt,Iyer:1994ys,Jacobson:1993vj,Dong:2013qoa}. The generalized volume $W_{\rm gen}$ was first conjectured in~\cite{Bueno:2016gnv}, which left the precise coefficients of various contributions undetermined. This expression is analogous to the original Wald entropy, which is derived for stationary event horizons on which the extrinsic curvature terms vanish. We fix the coefficients, as shown in eq.~\eqref{eq:details}, by carefully examining the higher-curvature corrections in eq.~\eqref{eq:CVsubregion2}. 
The term $W_K$ in eq.~\eqref{eq:ourporposal} generalizes the results to surfaces where the extrinsic curvature is non-vanishing, in analogy to Dong's extrinsic curvature corrections to the Wald entropy \cite{Dong:2013qoa}. These corrections naturally arise here in matching the subleading terms in the FG expansion of the volume of $\cal B$ in the bulk Einstein gravity case. However, as indicated in eq.~\eqref{eq:details}, we have only matched the corrections which are quadratic in $\mK_{\mu\nu}$ and as indicated by the ellipsis, this is only the first term in a longer expansion just as is found in the Wald-Dong entropy \cite{Dong:2013qoa}. We must also admit that even for the quadratic corrections, there is a high degree of ambiguity and the expression in eq.~\eqref{eq:details} is only the simplest ansatz consistent with our analysis.

The full analysis leading to these results is presented as follows:
In section \ref{sec:bulksubregion}, we exploit the Fefferman-Graham expansion near the brane to show that the leading-order contribution to holographic complexity coming from the island is given by the expression in eq.~\reef{eq:CVsubregion2}. In the process, we derive  the generalized complexity \reef{eq:ourporposal} for the effective higher-curvature theory of gravity on the brane. We also argue that the surface $\widetilde\mB$ on which the complexity is evaluated corresponds to the maximal complexity island. In section \ref{sec:GB}, we test our conclusions by beginning with a higher-curvature gravity theory in the ($d$+1)-dimensional bulk, \ie Gauss-Bonnet gravity and $f(\mR)$ gravity, and explicitly show our proposal \eqref{eq:ourporposal} consistently yields the same holographic complexity of islands as that derived from the effective gravitational theories on the brane. We present with a discussion of our results and future directions in section \ref{sec:dis}. In particular, we consider the quantum field theory corrections that implicitly appear when eq.~\reef{eq:CVsubregion2} is interpreted from the brane perspective. Appendix \ref{app:lower} contains some technical details that arise when studying the doubly holographic model in lower dimensions, \ie $d=2,3$.\\


%% file: sections/SubVolume.tex

In this section, we examine the subregion-CV conjecture in the context of the holographic model constructed in \cite{Chen:2020uac,Chen:2020hmv}. So we begin by reviewing some of the salient points of the model: As usual, the bulk gravity theory  is described by
\begin{equation}\label{eq:actionEinstein}
I_{\mt{bulk}}=\frac{1}{16 \pi \Gbulk} \int_{\rm bulk} d^{d+1} y \sqrt{-g}\left(\frac{d(d-1)}{L^{2}}+\mathcal{R}[g_{\mu\nu}]  \right) \,,
\end{equation}
where $L$ becomes the radius of curvature for the vacuum AdS$_{d+1}$ spacetime. Here, the bulk theory also includes a codimension-one brane with the action\footnote{We consider the addition of a DGP term below in section \ref{subsec:DGP}.}
\begin{equation}\label{eq:Ibrane}
I_{\rm brane}= -T_o \int d^d x \sqrt{-\tg} \,,
\end{equation}
where $T_o$ is the tension and $\tg_{ij}$ is the induced metric on the brane. 

\begin{figure}[t]
	\centering\includegraphics[width=5.8in]{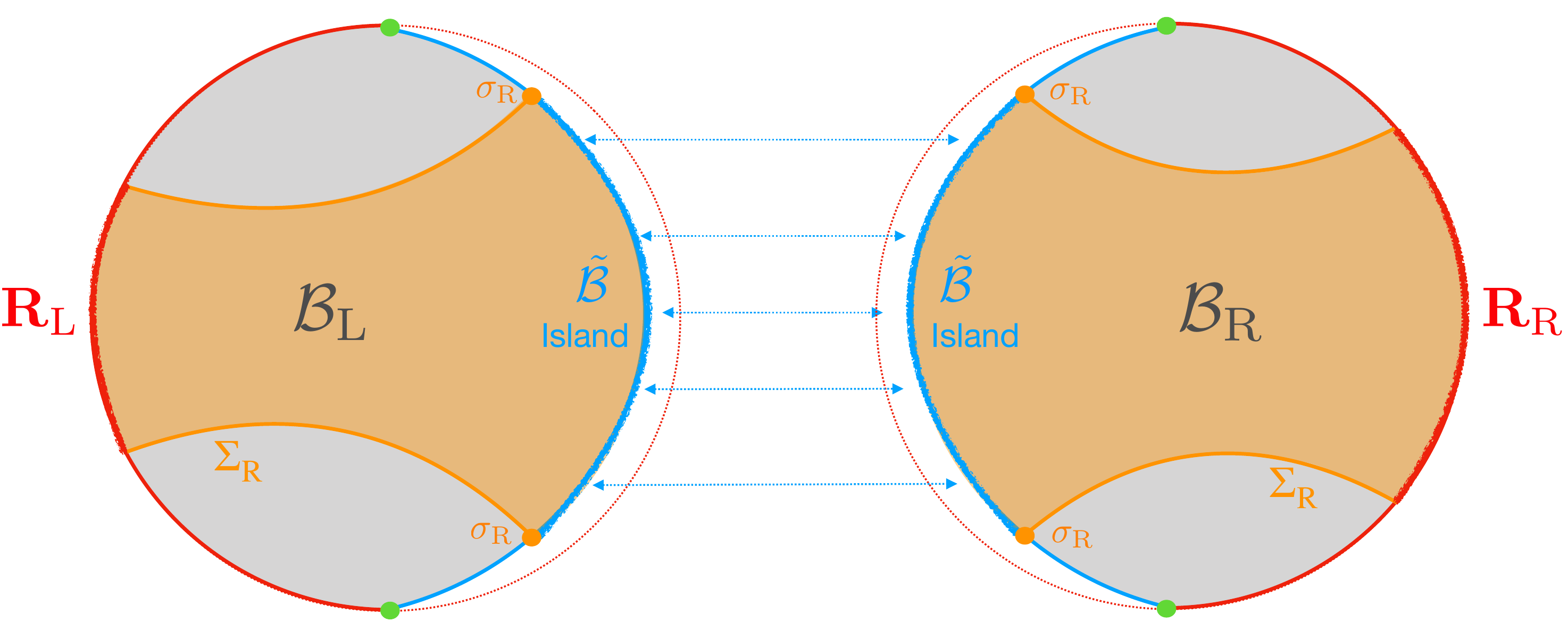}
	\caption{The holographic setup with islands in AdS$_{d+1}$. The two AdS$_{d+1}$ geometries are cut off at $\theta=\theta_\mt{B}$ (or $z=\zB$) and glued together with the brane at the junction between the two. The island region emerges on the brane when the RT surfaces $\Sigma_{\mathbf{R}}$ of the boundary subregion $\mA=\mA_{\mt{L}} \cup \mA_{\mt{R}}$ cross the brane. The maximal volume bulk slice $\mB = \mB_\mt{L} \cup \mB_\mt{R}$ crosses the brane, and the intersection of these two surfaces determines the island $\tilde{\cal B} = {\cal B} \cap {\rm brane} = {\cal B}_\mt{L} \cap {\cal B}_\mt{R}$.  \label{twoAdS}}
\end{figure}

Following \cite{Chen:2020uac,Chen:2020hmv}, we foliate of the bulk geometry with AdS$_d$ slices as in
\beq\label{metric33}
ds_{\ads{d+1}}^2
= \frac{L^2}{\sin^2\theta}\,\(d\theta^2 + ds^2_{\ads{d}}\)\,.
\eeq
where the AdS$_d$ metric has unit curvature. The solution with the brane is constructed by cutting the above geometry along an AdS$_d$ slice at some $\theta=\theta_\mt{B}$ near the asymptotic boundary. Joining together two copies of this geometry, as in figure \ref{twoAdS}, the brane is then represented as the interface between the two. That is, the brane is considered a shell of zero thickness and it's position the spacetime is determined using the Israel junction conditions \cite{israel1966singular} 
\begin{equation}\label{eq:junction}
\Delta (\mK_{\mt{B}})_{ij} - \tilde{g}_{ij} \Delta \mK_{\mt{B}}   =8 \pi \Gbulk\, S_{ij}=-8\pi \Gbulk\, T_o \,\tg_{ij}  \,, 
\end{equation}
where $S_{ij}$ is the boundary stress tensor introduced by the brane and $\Delta (\mK_{\mt{B}})_{ij} \equiv \mK^\mt{L}_{ij}- \mK^\mt{R}_{ij}$. The brane position can be written as
\beq
\sin^2\theta_\mt{B}=\frac{L^2}{\ell_\mt{B}^2}=
 2\,\veps\(1-\veps/2\)\qquad{\rm where}\ \ \veps\equiv\(1-\frac{4 \pi G_\mt{bulk} L T_o}{d-1}\)\,,
\label{curve2}
\eeq
and $\ell_\mt{B}$ is the curvature scale on the brane.

Now by construction, the bulk geometry locally takes the form of AdS$_{d+1}$ spacetime away from the brane. However, the brane's backreaction expands the bulk and with $\theta_\mt{B}\ll1$, the brane is pushed towards the asymptotic boundary of eq.~\reef{metric33}. Of course, this boundary (at $\theta=0$) is cut out of the construction, but we may still use the usual Fefferman-Graham (FG) expansion \cite{FG,Fefferman:2007rka} to examine the geometry in the vicinity of the brane. While the explicit construction described above is for the maximally symmetric ground state configuration, in the following, we consider more general configurations where the brane geometry may deviate slightly from the AdS$_d$ geometry. 

We begin by writing the metric on an asymptotically AdS$_{d+1}$ spacetime as\footnote{Our notation will be: Greek indices $\mu,\nu$ denote tensors in the bulk spacetime and run from $0$ to $d$. Latin indices $i,j$ from the middle of the alphabet denote tensors on codimension-one hypersurface at fixed $z$, and run from $0$ to $d-1$. For example, the bulk coordinates are $y^\mu=\lbrace z,x^i\rbrace$. Further, we will denote the bulk metric $g_{\mu\nu}= g^\mt{bulk}_{\mu\nu}$ in situations where there may be confusion. \label{convene}}
\begin{equation}\label{eq:FG}
\begin{split}
ds^2 = g_{\mu\nu}\, dy^\mu\, dy^\nu= \frac{L^2}{z^2} \(  dz^2 + g_{ij}(z,x^i) dx^i dx^j  \) \,.
\end{split}
\end{equation}
In these coordinates, we approach the asymptotic boundary for $z\to 0 $, and the brane is located at $z=\zB \ll L$.  Around the asymptotic boundary, the Fefferman-Graham expansion provides the a series expansion of the metric $g_{ij}(z,x^i)$ in terms of the boundary metric $\ov{0}{g}_{ij}$  and the boundary stress tensor $\ov{d/2}{g}_{ij}\propto \langle T_{ij}\rangle$ \cite{deHaro:2000vlm,Skenderis:2002wp}, \ie 
\begin{equation}\label{eq:bulkexpansion}
g_{ij}( z, x^i) =  \ov{0}{g}_{ij} \( x^i \)  + \frac{z^2}{L^2}\ov{1}{g}_{ij} \( x^i \) + \cdots  \frac{z^d}{L^d} \(  \ov{d/2}g_{ij} (x^i)  + f_{ij}(x^i) \log \( \frac{z}{L} \)    \) +\cdots\,,
\end{equation}
where the logarithmic term is present only when $d$ is even.
Now $\zB/L\ll1$ emerges as a natural expansion parameter, which we can apply in the FG expansion to study the geometry near the brane.

Applying the bulk Einstein equations in the FG expansion \reef{eq:bulkexpansion} fixes the expansion coefficients $\ov{n}{g}_{ij}$ (with $0<n < \frac{d}{2}$) in terms of the boundary metric $ \ov{0}{g}_{ij}$ \cite{deHaro:2000vlm,Skenderis:2002wp}. For example, the first term in the expansion is given by the Schoutten tensor $P_{ij}$ (for $d>2$),
\begin{equation}\label{eq:g1}
\ov{1}{g}_{ij} \( x^i \) = -L^2 P_{ij}[\ov{0}{g}] = -\frac{L^2}{d-2}  \( R_{ij}[\ov{0}{g}] - \frac{\ov{0}{g}_{ij} }{2(d-1)} R[\ov{0}{g}]   \) \,,
\end{equation}
where $R_{ij}$ and $R$ denote the Ricci tensor and Ricci scalar calculated with $\ov{0}{g}_{ij}$, respectively. We further note the above expression can also be derived by examining the effect of Penrose-Brown-Henneaux transformations \cite{Schwimmer:2008yh},  which implies that $\ov{1}{g}_{ij} \( x^i \) $ is completely determined by the conformal symmetries on the boundary and therefore it is independent of the bulk gravity theory. 
In contrast, the next term $\ov{2}{g}_{ij}$ in the expansion depends on the details of the bulk gravity theory, \eg see \cite{Imbimbo:1999bj,Myers:2013lva}. More precisely, it depends on whether the gravitational action contains interaction with the Riemann tensor squared, as we will see in section \ref{sec:GB}.

With the assumption that $\theta_\mt{B}\ll1$, one application of the  FG expansion \cite{FG,Fefferman:2007rka} is to derive the effective action for the gravity theory on the brane \cite{Chen:2020uac}
\beqa\label{eq:effectiveaction}
I_{\rm {eff}}&=&\frac{1}{16 \pi \Geff}  \int d^{d} x \sqrt{-\tilde{g}}\left[\frac{(d-1)(d-2)}{\ell_{\text {eff }}^{2}}+\tilde{R}(\tilde{g})\right] \\
&&\qquad+\frac{1}{16 \pi \Geff} \int d^{d} x \sqrt{-\tilde{g}}\left[\frac{L^{2}}{(d-4)(d-2)}\left(\tilde{R}^{i j} \tilde{R}_{i j}-\frac{d}{4(d-1)} \tilde{R}^{2}\right)+\cdots\right] \,,
\nonumber
\eeqa
where
\begin{equation}\label{eq:Gd}
\frac{1}{\Geff}=\frac{2L}{(d-2)G_{\rm bulk}}\,,
\qquad
\frac{1}{\ell_\mt{eff}^2}=
 \frac{2}{L^2}\,\veps \,,
\end{equation}
and $\tilde{g}_{ij}$ is the induced metric on the brane. The UV cutoff in this effective theory is given by $\tilde\delta=L$, and this controls the contributions of the higher curvature terms appearing in the second line of eq.~\reef{eq:effectiveaction}.\footnote{The ellipsis in eq.~\reef{eq:effectiveaction} indicates a further series of terms with higher powers of $L^2\,\times$ curvature.} Hence we are naturally lead to consider $\theta_\mt{B}\ll1$ (or equivalently, $L^2/\ell_\mt{eff}^2\ll1$ or $\veps\ll1$) as this corresponds to the regime in which the induced brane theory is well approximated by Einstein gravity with a negative cosmological constant. 

Similarly, the FG expansion can be applied to understand the contributions of the holographic entanglement entropy \reef{eq:island2} in terms of the brane theory, \eg one finds that the gravitational contribution in the island rule \reef{eq:islandformula} corresponds to the Wald-Dong entropy for the induced action \reef{eq:effectiveaction} evaluated on the boundaries of the island \cite{Chen:2020uac}. In the following, we follow a similar strategy applying the FG expansion to examine the bulk holographic complexity \reef{eq:CVsubregion} evaluated in the vicinity of the brane and reinterpret the result in terms of the brane theory. In particular, we will find the geometric contributions in the `island' complexity, and provide a prescription to derive these from the effective action \reef{eq:effectiveaction}.

\begin{figure}[t]
	\centering\includegraphics[width=5.8in]{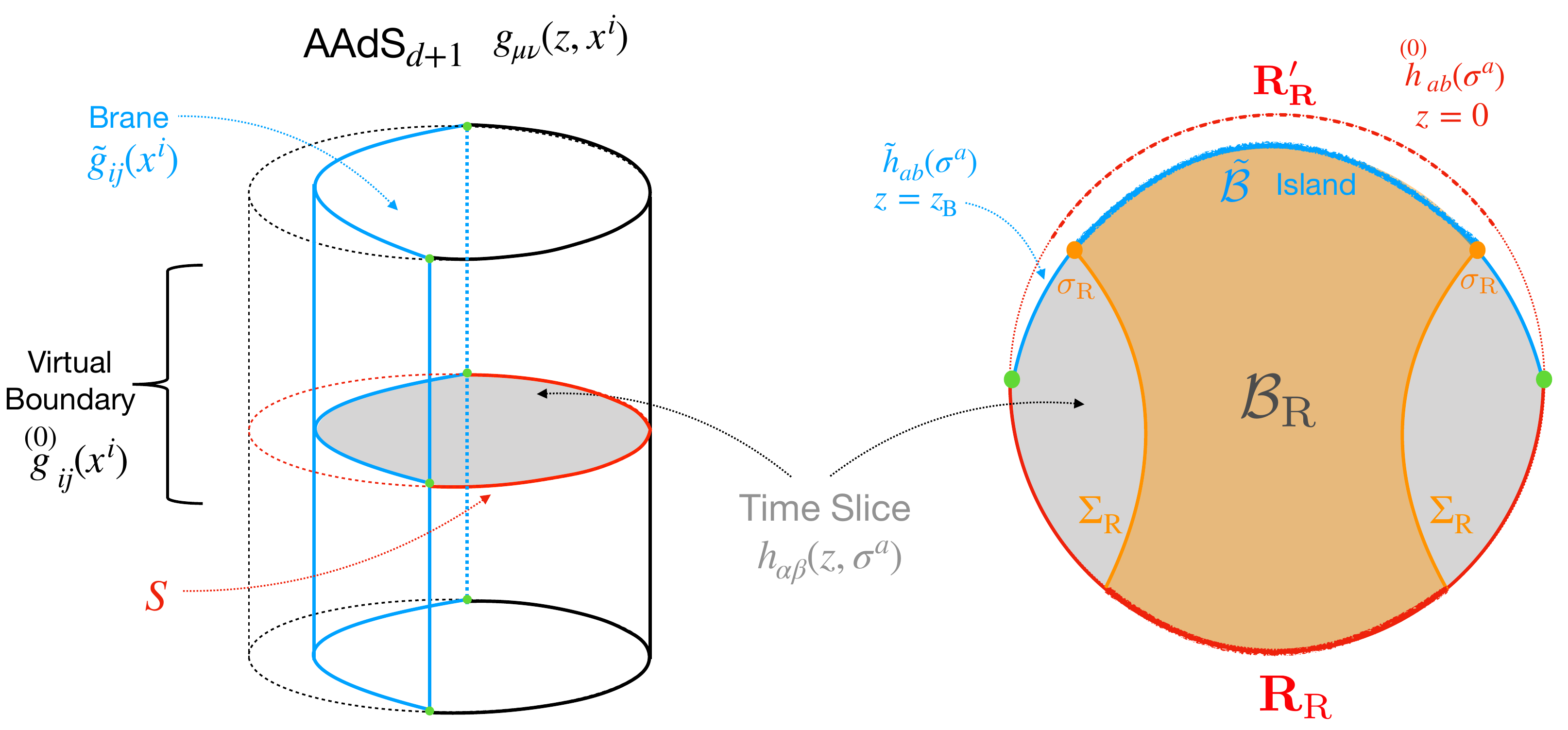}
	\caption{The full asymptotically AdS$_{d+1}$ geometry from the right side of the construction in figure \ref{twoAdS}. The time slice $\mS$ is introduced in the left panel and detailed in the right panel. We explicitly show various metrics for the different regions.  \label{AdSCylidner}}
\end{figure}

\subsection{Extremal surfaces near the brane}\label{subsec:CVEinstein}

Eq.~\reef{eq:CVsubregion} gives the complexity=volume proposal for a boundary subregion $\mA$  as,
\begin{equation}
\label{eq:CV-notboth}
{\cal C}_{\mt{V}}^{\rm{sub}} \(\mA\) = \max_{\partial {\cal B}=\mA\cup \RT}  \[ \frac{V\(\mB\)}{\Gbulk\, \ell}\] \,.
\end{equation}
In particular, one extremizes the volume of codimension-one hypersurface ${\cal B}$ anchored on the subregion $\mA$ on the asymptotic boundary and on the RT surface $\RT$ in the bulk. 
Since we are interested in reinterpreting the bulk results in terms of the brane theory, we will assume that we are in the island phase, \ie the RT surface $\RT$ crosses the brane, as shown in figure \ref{twoAdS}. Then, as shown, our boundary subregion $\mA$ will generally have components $\mA_\mt{L}$ and $\mA_\mt{R}$ on either side of the conformal defect in the boundary theory. Similarly, we decompose the bulk surface in terms of components on either side of the brane, \ie $\mB=\mB_\mt{L} \cup \mB_\mt{R}$. We also remark that in applying the FG expansion, we extend the left or right geometry to a `virtual' asymptotic boundary at $z=0$, so that the `boundary metric' $\ov{0}{h}_{ab}$ and other boundary quantities are evaluated at the region $\mA'_{\mt{R}}$ (and similarly a region $\mA'_{\mt{L}}$ for the left AdS region) at this virtual boundary, as shown in the right panel of figure~\ref{AdSCylidner}.

To facilitate our analysis, we introduce $d$-dimensional coordinates $\sigma^\alpha$ in $\cal B$ with letters from the beginning of the Greek alphabet, \ie $\alpha, \beta, \gamma$ which run from $1$ to $d$. Further, we use Gaussian normal coordinates with respect to the intersection $\wmB = {\cal B}\cap {\rm brane}$, with $\zeta=\sigma^d$ being the coordinate normal to the brane. Latin indices $a, b, c$ from the beginning of the alphabet denote the other directions running from $1$ to $d -1$, \ie $\sigma^\alpha= (\zeta,\sigma^a)$.s Taking the parametrization of the bulk hypersurface $\mB$ as $y^\mu \( \zeta, \sigma^a \)$, we can define the induced metric on this surface by
\begin{equation}
h_{\alpha\beta} = \frac{\partial y^\mu}{\partial \sigma^\alpha}\,\frac{\partial y^\nu}{\partial \sigma^\beta}\, g_{\mu\nu}[y]\,.
\end{equation}
As a bulk tensor, we may also write the induced metric as
\beq
h_{\mu\nu} = \[g_{\mu\nu}\]_\mB + n_{\mu}n_{\nu}\,,
\eeq
where $n^\mu$ is the unit vector normal to $\cal B$, \ie $n^\nu n^\mu g_{\mu\nu} =-1$ and $h_{\mu\nu} n^\nu=0$. Further, it will be convenient to make the following gauge choices:
\begin{equation}
\zeta=\sigma^d =z\qquad{\rm and} \qquad h_{z a}= 0\,.
\end{equation}

In order to consider holographic complexity for $(d+1)$-dimensional bulk theory, we are interested in the codimension-1 bulk surface $\mathcal{B}$ with extremal volume in the bulk. 
Extremizing the volume of hypersurface $\mathcal{B}$ leads to 
a local equation 
\begin{equation}\label{eq:CVmaximal}
\text{EOM}^\mu=\frac{1}{\sqrt{h}}\,\partial_\alpha\! \(  \sqrt{h} h^{\alpha\beta} \partial_\beta y^\mu \)  + h^{\alpha\beta} \partial_\alpha y^\nu\partial_\beta y^\sigma \,\Gamma^\mu_{\nu\sigma}=0\,,
\end{equation}
where $h=\det h_{\alpha\beta}$ and $\Gamma^\mu_{\nu\sigma}$ is the Christoffel symbol associated with the bulk metric $g_{\mu\nu}$.  As a vector, the above expression is orthogonal to $\mB$ and taking the inner product with $n^\mu$ leaves a simple expression in terms of the extrinsic curvature $\mK_{\alpha\beta}$ of the submanifold (see eq.~\eqref{eq:Gauss}),
\begin{equation}\label{eq:extremalEinstein}
\mK  = h^{\alpha\beta}\,\mK_{\alpha \beta} = 0 \,.
\end{equation} 

Since we are interested in the geometry near the asymptotic boundary, above equation can be solved order by order in a Fefferman-Graham expansion for $x^i\( z, \sigma^a \)$
\begin{equation}\label{eqx1FG}
x^i \( z, \sigma^a \) = \ov{0}{x^i}\(\sigma^a \)  + \frac{z^2}{L^2}  \,\ov{1}{x^i} \( \sigma^a\)+ \mathcal{O}\(\frac{z^4}{L^4}\)\,. 
\end{equation}
Noting that the leading contribution in eq.~\eqref{eq:CVmaximal} involves the terms with two $z$ derivatives,  we see that the extremization condition does not fix the leading coefficients $\ov{0}{x}{}^i$, \ie the profile of the surface at $z=0$. Alternatively, we can think of this indeterminacy as the profile of the intersection of the extremal surface $\cal B$ and the brane, which we will refer to as the island $\wmB=\mB\,\cap\,$brane. As we will emphasize in section~\ref{subsec:maximal}, solving eq.~\eqref{eq:CVmaximal} or \reef{eq:extremalEinstein} ensures that the volume of $\cal B$ is extremized in the bulk, \ie away from the brane. Producing the correct maximal volume surface in eq.~\reef{eq:CV-notboth} requires a second step where we vary the island profile $\wmB$ which maximizes complexity functional on the brane -- see eqs.~\eqref{eq:new} and \eqref{eq:CVequalsisland}. 

Following the analysis in, \eg \cite{Schwimmer:2008yh,Carmi:2016wjl,Chen:2020uac}, 
the leading order terms in eq.~\eqref{eq:CVmaximal} are
\begin{equation}\label{eq:exapndEOM}
\frac{2z^2}{L^2}(1-d) \ov{1}{x^i}  + \frac{1}{\sqrt{\ov{0}{h}}} \partial_a\( \sqrt{\ov{0}{h}} \ov{0}{h^{ab}}\partial_b x^i\) + \ov{0}{h^{ab}}\Gamma^i_{ik} \partial_a \ov{0}{x^j} \partial_b \ov{0}{x^k}+ \mathcal{O}(z^4) =0\,.
\end{equation}
 Thus the the first order term in the FG expansion for $x^i$ is given by 
\begin{equation}\label{eq:X1}
\ov{1}{x^i} \( \sigma^a\)= \frac{L^2}{2(d-1)} \(  \ov{0}{D^a}( \partial_a \ov{0}{x^i})  + \ov{0}{h^{ab}} \partial_a \ov{0}{x^j} \partial_b \ov{0}{x^j} \Gamma^i _{jk}\) = \frac{L^2}{2(d-1)} K \ov{0}{n^i}  \,,
\end{equation}
where $\ov{0}{D_a}$ denotes the covariant derivative associated with induced metric $\ov{0}{h_{ab}}$ on the (implicit) boundary time slice at $z=0$,  $K$ is the trace of extrinsic curvature for this time slice (\ie $K=\ov{0}{g^{ij}}K_{ij}$), and $\ov{0}{n^i}$ denotes the timelike unit normal to the same time slice (\ie $\ov{0}{n^i}\ov{0}{ n^i} \ov{0}{g}_{ij}=-1$). In order to get the second equality in eq.~\eqref{eq:X1}, we have used the trace of Gauss-Weingarten equation, which reads 
\begin{equation}\label{eq:Gauss}
e_b^j \nabla_j (e^i_a) = \Gamma^c_{ab} e^i_c + \, K_{ab} n^i \,,
\end{equation}
after taking $e_a^i \equiv \partial_a \ov{0}{x^i}$. 
The above result is very similar to the solutions for the extremal RT surface in a ($d+2$)-dimensional bulk model, although in this case, we are working with a codimension-one hypersurface. With the asymptotic solutions,  we find the induced metric components on the extremal surface $\mB$ read 
\begin{equation}
\begin{split}
h_{zz}&= \frac{L^2}{z^2} \( 1+ \frac{z^2}{L^2}\frac{4\ov{1}{x^i}\ov{1}{x^j}}{L^2}\ov{0}{g}_{ij}   +\cdots \)=\frac{L^2}{z^2} \( 1-\frac{z^2}{(d-1)^2}\, K^2   + \cdots  \)  \,,\\
h_{ab}&=\frac{L^2}{z^2} \( \ov{0}{h}_{ab} + \frac{z^2}{L^2}\ov{1}{h}_{ab}   +\cdots\) \,,\\
\end{split}
\end{equation}
with 
\begin{equation}
\begin{split}
\ov{0}{h}_{ab}   =\ov{0}{g}_{ij}\, \partial_a \ov{0}{x^i} \partial_b \ov{0}{x^i} \,, \quad \ov{1}{h_{ab}}= \ov{1}{g_{ab}} + \frac{L^2}{d-1} K K_{ab}\,,
\end{split}
\end{equation}
where the tensors with indices $a,b$ are associated with those with $i,j$ by using the projection $\partial_a \ov{0}{x^i} \equiv  e_a^i$.

Following the subregion-CV proposal \reef{eq:CV-notboth}, our goal is to find the maximal volume hypersuface $\mB$ anchored on the boundary subregion $\mA$ and the bulk RT surface $\RT$, \ie $\partial {\cal B} = \mA\cup \RT$, and then evaluate
\begin{equation}
\mC^{\rm{sub}}_\mt{V} (\mA) = \frac{V (\mB)}{\Gbulk\ell}= \frac{1}{\Gbulk \ell} \int_{\mathcal{B}} d^{d-1}\sigma d z\, \sqrt{\det h_{\alpha\beta}}\,.
\end{equation}
In the present calculation with the brane positioned at $\zB\ll L$, we are particularly interested the contributions to the maximal volume coming from the region in the vicinity of the brane.\footnote{Note that we ignore here the UV-divergent contributions coming from the asymptotic boundary \cite{Carmi:2016wjl}. These are less interesting for our purposes and might be eliminated by considering the mutual complexity \cite{Alishahiha:2018lfv,Caceres:2019pgf} -- see the discussion section.} Approaching $z\to0$, the volume measure reduces to 
\begin{equation}
\sqrt{\det h_{\alpha\beta}} = \sqrt{\det \ov{0}{h}_{ab}} \( \frac{L}{z} \)^d\( 1-\frac{z^2}{2(d-1)^2} K^2 + \frac{z^2}{2L^2} \ov{0}{h^{ab}}\ov{1}{h_{ab}} + \cdots \)\,. 
\end{equation}
where we have ignored the contributions from higher order $\zB/L$ terms. Performing the $z$-integral explicitly and introducing $R^a_a[\ov{0}{g}] = \ov{0}{h^{ab}} R_{ab}[\ov{0}{g}]$, we can find the leading contributions of the holographic subregion-complexity near the brane
\begin{equation}\label{eq:Vbulk}
\frac{L^d}{\Gbulk \ell} \int_{\wmB} d^{d-1}\sigma \,\sqrt{\det \ov{0}{h}_{ab}}  \left[ \frac{1}{(d-1)\zB^{d-1}} + \frac{1}{(d-3)\zB^{d-3}} \(  \frac{d-2}{2(d-1)^2} K^2  -\frac{R^a_a- \frac{1}{2}R}{2(d-2)} \) +\cdots \right]\,,
\end{equation}
where the extrinsic curvature and Ricci tensor are all related to boundary geometry at $z=0$. 

We can also evaluate the volume of the island region 
\begin{equation}\label{eq:Visland}
\begin{split}
V(\wmB) &= \int_{\wmB} d^{d-1}\sigma \, \sqrt{\det \tilde{h}_{ab}}\,, \\
&=L^{d-1} \int_{\wmB} d^{d-1}\sigma \,\sqrt{\det \ov{0}{h}_{ab}}  \left[ \frac{1}{z_{\mt{B}}^{d-1}} + \frac{1}{z_{\mt{B}}^{d-3}} \(  \frac{K^2}{2(d-1)}  -\frac{R^a_a- \frac{1}{2}R}{2(d-2)} \) +\cdots  \right]\,. 
\end{split}
\end{equation}
with 
$\tilde{h}_{ab}\equiv h_{ab}(z=z_{\mt{B}})$ 
as the induced metric on the intersection 
$\widetilde\mB=\mB\,\cap\,$brane. 
Combing eqs.~\eqref{eq:Vbulk} and \eqref{eq:Visland}, it is straightforward to rewrite the holographic subregion-complexity \reef{eq:Vbulk} as
\begin{equation}\label{eq:full_subCV}
\begin{split}
&\mC^{\rm{sub}}_\mt{V} \( \mA\) = \frac{V\( \mathcal{B}\)}{\Gbulk \ell} \\
&\simeq \frac{2L V(\wmB) }{(d-1)\Gbulk \ell} +  \frac{2}{\Gbulk \ell}\int_{\wmB} d^{d-1}\sigma \,\sqrt{\det \ov{0}{h}_{ab}}   \frac{L^d}{z_{\mt{B}}^{d-3}} \(  \frac{K^2}{2(d-1)^2(d-3)}  -\frac{R^a_a- \frac{1}{2}R}{(d-1)(d-2)(d-3)} \) +\cdots \\
&\simeq \frac{2L V(\wmB) }{(d-1)\Gbulk \ell} +  \frac{2L^3}{\Gbulk \ell}\int_{\wmB} d^{d-1}\sigma \,\sqrt{\det \tilde{h}_{ab}}   \(  \frac{\tilde{K}^2}{2(d-1)^2(d-3)}  -\frac{\tilde{R}_{ij} \tilde{n}^i \tilde{n}^j + \frac{1}{2}\tilde{R}}{(d-1)(d-2)(d-3)} \) +\cdots \,. 
\end{split}
\end{equation}
where the factor of $2$ above originates from the fact that we are integrating over both sides of the island, \ie we are including the contributions from both $\mathcal{B}_{\mt{L}}$ and $\mathcal{B}_{\mt{R}}$. Furthermore, we note that we {\it {do not}} need to require a symmetric setup\footnote{That is, we do not require an $\mathbb{Z}_2$ symmetry about the brane, as was imposed in the explicit calculations performed in \cite{Chen:2020uac,Chen:2020hmv}.} because the near-brane regions from $\mathcal{B}_{\mt{L}}, \mathcal{B}_{\mt{R}}$ have the same leading order contributions, despite the fact that the full volume of the subregions $\mathcal{B}_{\mt{L}}, \mathcal{B}_{\mt{R}}$ may be different. Of course, while the surfaces $\mB_\mt{L}$ and $\mB_\mt{R}$ are independent away from the brane, their profiles on the brane coincide, \ie $\wmB = B_{\rm L} \cap B_{\rm R}$. Let us also note here that $\wmB$ is anchored to the intersection of the RT surface $\RT$ with the brane, \ie $\partial\wmB=\RTbrn=\RT\,\cap\,$brane, but this is precisely the quantum extremal surface (QES) in the brane theory \cite{Chen:2020uac,Chen:2020hmv}.

To arrive at the last line of eq.~\eqref{eq:full_subCV}, we recast the boundary terms into terms related to the brane geometry following \cite{Chen:2020uac}. First we note that the induced metric on the brane reads 
\begin{equation}\label{eq:rescale}
\begin{split}
\tilde{g}_{ij}(x^i) &\equiv g^\mt{bulk}_{ij} (z_{\mt{B}}, x^i)=\frac{L^2}{z^2_{\mt{B}}} \,g_{ij}(z_\mt{B},x^i) \approx \frac{L^2}{z^2_{\mt{B}}} \,\ov{0}{g_{ij}}(x^i) + \mathcal{O}\(z^0_{\mt{B}}\) \,,\\
\tilde{h}_{ab}&\equiv h_{ab}(z=z_{\mt{B}}) \approx \frac{L^2}{z^2_{\mt{B}}} \,\ov{0}{h_{ij}}(x^i) + \mathcal{O}\(z^0_{\mt{B}}\) \,,\\
\end{split}
\end{equation}
as well as using $\tilde{h}_{ij}= \tilde{g}_{ij}+ \tn_i \tn_j$, where $\tn^i$ denotes the unit time-like normal to island in the brane. We therefore find 
\begin{equation}
\begin{split}
\frac{\zB^2}{L^2}\ov{0}{h^{ab}} e^i_ae^j_b \( R_{ij}[\ov{0}{g}] -  \frac{\ov{0}{g}_{ij} }{2(d-1)} \mathcal{R}[\ov{0}{g}]   \) &\approx {\tilde{h}^{ab}} \( \tilde{R}_{ab}[\tilde{g}] -  \frac{ \tilde{h}_{ab} }{2(d-1)} \tilde{R}[\tilde{g}]    \)  \\
&\approx \frac{1}{2}\tilde{R}[\tilde{g}]+ \tilde{R}_{ij}[\tilde{g}]\tn^i\tn^j \,,
\end{split}
\end{equation}
by keeping track of the leading contributions in the $z_{\mt{B}}/L$ expansion. As expected, the leading term in $\mC^{\rm{sub}}_\mt{V} \( \mA\)$ is the volume of island region. Interestingly, the result in eq.~\eqref{eq:full_subCV} shows that the subleading terms include intrinsic geometric quantities on the brane but also include the extrinsic curvature of the island region $\wmB$, \ie the term proportional to $\tK^2$. This feature is also found in a similar analysis for holographic entanglement entropy in section 4.3 of~\cite{Chen:2020uac}. 

Now examining eq.~\reef{eq:full_subCV}, we see to leading order that we have
\beq\label{scrabble0}
\mC^{\rm{sub}}_\mt{V} \( \mA\) = \frac{V\( \mathcal{B}\)}{\Gbulk \ell} = \frac{2L\, V(\wmB) }{(d-1)\Gbulk \ell} +\cdots =\frac{d-2}{d-1} \,\frac{ V(\wmB)}{\Geff\, \ell} + \cdots  \,, \\
\eeq
where $\frac{1}{\Geff}= \frac{2L}{(d-2)\Gbulk}$ is the effective Newton's constant for the brane gravity,  as given in eq.~\reef{eq:Gd}. That is, the complexity=volume formula in the bulk yields a complexity=volume formula on the brane, up to an inconvenient numerical factor. Now this factor could be easily absorbed if we modify the length scale for the CV proposal on the brane, \ie
\beq\label{newel}
\ell'=\frac{d-1}{d-2}\,\ell\,.
\eeq
However, beyond the volume term, eq.~\reef{eq:full_subCV} also contains higher-order corrections involving the curvature on the brane and the extrinsic curvature of the surface $\wmB$. By examining these contributions more carefully in the next subsection, we will be able to interpret them in terms of a generalized CV formula derived from the induced higher-curvature gravity action \reef{eq:effectiveaction} on the brane. The emergence of this generalized CV expression in the brane theory is then analogous to the appearance of the Wald-Dong entropy in the island rule \reef{eq:islandformula} on the brane discussed in \cite{Chen:2020uac}.

\subsection{Holographic complexity on the brane} \label{OTbrain}

In this subsection, we show that the sub-leading contributions in eq.~\eqref{eq:full_subCV} can be consistently derived from the induced gravity action in eq.~\reef{eq:effectiveaction} with a simple generalization of the complexity=volume prescription in eq.~\reef{eq:CV-notboth}. The question of extending the CV proposal to higher curvature theories of gravity was first considered in \cite{Bueno:2016gnv}. For a gravitational theory in $d+1$ dimensions, their proposal was that the usual volume functional should be replaced by a generalized volume of the following form
\begin{equation}\label{eq:defineWgen}
{W}_{\rm gen}\!\({\mB}\) = \int_{\mB}  d^{d}\sigma\,  \sqrt{{h}} \(   \frac{\partial \mathbf{L}}{\partial R_{ijkl}}{h}_{jk} \(  \alpha_{d+1} n_i n_l + \beta_{d+1} {h}_{il} \) + \gamma_{d+1} \)
\end{equation}
where $\alpha_{d+1}$, $\beta_{d+1}$ and $\gamma_{d+1}$ are numerical constants (depending on the boundary dimension $d$).

However, this suggestion by itself can not provide the extrinsic curvature terms in eq.~\eqref{eq:full_subCV}. A similar issue was encountered in extending holographic entanglement entropy to higher curvature theories. In particular, it was shown that replacing the Bekenstein-Hawking entropy with the Wald entropy~\cite{Wald:1993nt,Iyer:1994ys,Jacobson:1993vj} in the RT prescription will not produce the expected entanglement entropy for the boundary theory \cite{Hung:2011xb}. Instead, the correct extension required the addition of `corrections' involving the extrinsic curvature of the extremal surface in the bulk  \cite{Dong:2013qoa}. Hence we propose the generalized CV prescription for higher curvature gravity theories must include additional $K$-terms. Explicitly, we suggest that the leading contributions take the form
\begin{equation}
\begin{split}
\label{eq:defineWK}
W_K\!\({\mB}\)=\int_{\mB} d^{d}\sigma\,  \sqrt{{h}}\, \bigg[ \frac{\partial^2 \mathbf{L}}{\partial R_{ijkl}\partial R^{mnop}}&\, K_{jl}\(  A_{d+1} h_{ik} +  B_{d+1} n_i n_k\)\\
\times\ \ & K^{np} \( A_{d+1} h^{mo} + B_{d+1} n^m n^o \) \bigg]\,,
\end{split}
\end{equation}
where again $A_{d+1}$ and $B_{d+1}$ are numerical constants. 

Correspondingly, we propose that the holographic complexity for the island region on the brane can be derived from
\begin{equation}\label{eq:new}
\mC_\mt{V}^{\rm Island}=\max_{\partial\wmB=\RTbrn} \[  \frac{\widetilde{W}_{\rm gen}(\wmB) + \widetilde{W}_K(\wmB)}{\Geff\,\ell'}  \] \,,
\end{equation}
where $\RTbrn=\RT\,\cap\,$brane is the quantum extremal surface on the brane -- see figure \ref{twoAdS}. We have introduced the
notation $\widetilde{W}_{\rm gen},\, \widetilde{W}_K$ to indicate these are quantities defined for the $d$-dimensional gravity theory on the brane. In the following subsections, we seek to compare $\mC_\mt{V}^{\rm Island}$ with the leading terms in the holographic CV found in eq.~\eqref{eq:full_subCV} to fix the numerical coefficients in eqs.~\reef{eq:defineWgen} and \reef{eq:defineWK}. This proposal also requires that we maximize the new functional over all profiles $\wmB$ anchored to  the QES $\RTbrn$, but we leave the discussion of this point to section \ref{subsec:maximal}. 

\subsubsection{Generalized Volume on the Island}

Substituting eq.~\reef{eq:Gd} for effective Newton's constant and eq.~\reef{newel} for the CV length scale on the brane into the last line of eq.~\eqref{eq:full_subCV}, the leading contribution to the holographic complexity becomes
 \beqa\label{eq:CV_full}
 \mC^{\rm{sub}}_\mt{V} \( \mA\)&=& \frac{V(\wmB)}{\Geff\, \ell'} 
 \\
 &&\quad +\ 	\frac{L^2}{\Geff\, \ell'}\int_{\wmB} d^{d-1}\sigma \,\sqrt{\tilde{h}}\(  \frac{\tilde{K}^2}{2(d-1)(d-3)}  -\frac{\frac{1}{2}\tilde{R}[\tilde{g}]+ \tilde{R}_{ij}[\tilde{g}]\tilde{n}^i\tilde{n}^j}{(d-2)(d-3)}  +\cdots \)\,.
 \nonumber 
 \eeqa
Now our aim is to show that these results can be derived from our proposal for the complexity of the island in eq.~\eqref{eq:new} applied to the effective gravitational action \reef{eq:effectiveaction}. In particular, to make this match, we must choose the appropriate numerical constants $\alpha_d$, $\beta_d$, $\gamma_d$, $A_d$ and $B_d$ for the $d$-dimensional brane theory. Here, we focus on the first three coefficients appearing in the generalized volume $\widetilde{W}_{\rm gen}(\wmB)$.

To begin with, we consider a general quadratic Lagrangian as
\begin{equation}
\mathbf{L}_{\rm eff} \equiv 16\pi \Geff\mathcal{L} = \tilde{R} -2\Lambda + \lambda_1 \tilde{R}^2 + \lambda_2 \tilde{R}_{ij} \tilde{R}^{ij}   \,. 
\end{equation}
We want to evaluate the generalized volume for the complexity $\mC_{\mt{V}}^{\rm Island}$ in eq.~\eqref{eq:new}. 
Using the Kronecker delta of rank-two (\ie $\delta^{ij}_{mn}= \delta^i_m\delta^j_n - \delta^i_n\delta^j_m$),  the derivative with respect to Riemannian tensor is explicitly written as \cite{Miao:2013nfa}  
\begin{equation}\label{eq:firstderivative}
\frac{\partial \tR_{mnop}}{\partial \tR_{ijkl}} \equiv (\partial \tR)^{ijkl}_{mnop} =\frac{1}{12} \( \delta^{ij}_{mn} \delta^{kl}_{op}  -\frac{1}{2} \delta^{ik}_{mn} \delta^{lj}_{op}  -\frac{1}{2} \delta^{il}_{mn} \delta^{jk}_{op} + \delta^{ij}_{op} \delta^{kl}_{mn}  -\frac{1}{2}\delta^{ik}_{op} \delta^{lj}_{mn}  -\frac{1}{2}\delta^{il}_{op} \delta^{jk}_{mn} \).
\end{equation}
It is then straightforward to get the tensor
\begin{equation}
\begin{split}
\frac{\partial \mathbf{L}_{\rm eff}}{\partial R_{ijkl}}&=\(  \frac{1}{2}   + \lambda_1 \tR\) 2 \tg^{i[k}\tg^{l]j} +  \lambda_2 \( \tR^{i[k} \tg^{l]j} +\tR^{j[l} \tg^{k]i}  \)\,, \\
\end{split}
\end{equation}
where $Z^{[ij]}=\frac{1}{2}\(Z^{ij}-Z^{ji} \) $. One can explicitly evaluate the needed contractions to find 
\beqa
&& \frac{\partial \mathbf{L}_{\rm eff}}{\partial R_{ijkl}}\tilde{h}_{jk} \(  \alpha_d \tn_i \tn_l + \beta_d \tilde{h}_{il} \) + \gamma_d  = \gamma_d +\(  \frac{1}{2}   + \lambda_1 \tR\) (d-1) \(\alpha_d - (d-2) \beta_d  \) \nonumber\\
&&\qquad\qquad+\frac{\lambda_2}{2} \( \tR (\alpha_d - 2(d-2) \beta_d) - \tR^{ij}\tn_i \tn_j (\alpha_d +2\beta_d )(d-2) \)  \,.
\label{eq:Leff}
\eeqa
Comparing the above results with eq.~\eqref{eq:CV_full}, 
and taking the effective action~\eqref{eq:effectiveaction} on the brane, \ie choosing the two coupling constants as 
\begin{equation}
\lambda_1 =-\frac{dL^2}{4(d-1)(d-2)(d-4)}  \,,\qquad \lambda_2= \frac{L^2}{(d-2)(d-4)} \,,
\end{equation}
one finds that the three coefficients in the generalized volume should be fixed to  
\begin{equation}\label{eq:alphagamma}
\alpha_d = \frac{2 (d-4)}{(d-2)(d-3)} \,, \quad  \beta_d =0 \,, \quad \gamma_d= \frac{2 }{(d-2)(d-3)}  \,.
\end{equation}

As a recap, the comparison between the leading contributions to the volume of the extremal surface $\mB$ in the vicinity of the brane for $(d+1)$-dimensional bulk gravity theory in eq.~\eqref{eq:CV_full} and the generalized volume on the brane determines the numerical coefficients in the latter as in eq.~\eqref{eq:alphagamma}. Hence, the resulting generalized volume reads 
\begin{equation}\label{eq:fixW}
\widetilde{W}_{\rm gen} ( \wmB )= \frac{2 }{(d-2)(d-3)}\  \int_{\wmB } d^{d-1}\sigma \,\sqrt{\det \tilde{h}_{ab}} \left(1+(d-4)\, \frac{\partial \mathbf{L}_{\rm eff}}{\partial \tR_{ijkl}} \tn_{i} \tilde{h}_{jk} \tn_{l} \right) \,.\\
\end{equation}
Furthermore, we propose that this result of the generalized volume can be used in extending the holographic complexity=volume conjecture for higher curvature gravity theories in general, as in eqs.~\reef{eq:ourporposal} and \reef{eq:details}. In section~\ref{sec:GB}, we will test this proposal further by considering higher curvature gravity in the bulk of our holographic model.
 
\subsubsection{$K$-term on the Island} \label{klumbsy}

As discussed above, the generalized volume \eqref{eq:defineWgen} by itself fails to provide the full holographic complexity on the island due to the appearance of terms involving the extrinsic curvature $\tK$ on the brane. Inspired by the Wald-Dong entropy, we suggested the addition of $K$-terms to the generalized volume. At the second-order, we can produce a covariant quantity by contracting the tensor $\frac{\partial^2 \mathbf{L}_{\rm eff}}{\partial \tR_{ijkl} \partial \tR^{mnop}}$, with the tensors built from the three independent symmetric tensors $\tK_{ij}, \tilde{h}_{ij}, \tn_i \tn_j$.  The simplest choice is the following 
\begin{equation}\label{scrabble4}
\widetilde{W}_{K}( \wmB)=\int_{\wmB} d^{d-1}\sigma \,\sqrt{\tilde{h}}\, \frac{\partial^2 \mathbf{L}_{\rm eff}}{\partial \tR_{ijkl}\partial \tR^{mnop}} \tK_{jl}\(  A_d \tdh_{ik} +  B_d \tn_i \tn_k\) \tK^{np} \( A_d \tdh^{mo} + B_d \tn^m \tn^o \) \,,
\end{equation}
where as before, $\mathbf{L}_{\rm eff}=16 \pi \Geff\mathcal{L}_{\rm eff}$. Our goal is then to fix the two numerical coefficients $A_d, \,B_d$. 

To compute $\frac{\partial^2 \mathbf{L}_{\rm eff}}{\partial \tR_{ijkl} \partial \tR^{mnop}}$, we  need to use the second derivative  
\beqa\label{eq:secondderivative}
\frac{\partial^2 (\tR^2)}{\partial \tR_{ijkl} \partial \tR^{mnop}} &=&\frac{1}{2}\( \tg^{ik}\tg^{jl} - \tg^{il}\tg^{jk}  \) \(  \tg_{mo}\tg_{np} -\tg_{mp}\tg_{no}   \)\,,\\
\frac{\partial^2 (\tR_{i_1j_1}\tR^{i_1j_1})}{\partial \tR_{ijkl} \partial \tR^{mnop}} &=&\frac{1}{2}\tg_{rs}  \(  (\partial \tR)^{irks}_{mnop} \, \tg^{jl}   -(\partial \tR)_{mnop}^{irls} \, \tg^{jk} -(\partial \tR)_{mnop}^{jrks} \, \tg^{il} +(\partial \tR)_{mnop}^{jrls}\, \tg^{ik}  \)  \,,
\nonumber
\eeqa
where the tensor $(\partial \tR)^{ijkl}_{mnop} $ is the first derivative defined in eq.~\eqref{eq:firstderivative}. 

Applying the second derivative \eqref{eq:secondderivative} to the effective action in eq.~\eqref{eq:effectiveaction},
one finds that the proposed $\widetilde{W}_K$ reduces to 
\beqa\label{eq:effectiveWK}
 \widetilde{W}_{K}( \wmB)&=& \int_{\wmB} d^{d-1}\sigma \,\sqrt{\tilde{h}}\, \bigg[ \frac{\lambda_1\tK^2}{2} \(  (d-2)A_{d} -B_{d} \)^2   \\
&&\quad + \frac{\lambda_2}{8} \( \tK^2\(B_{d}^2 -2A_{d}B_{d} +(3d-7)A_{d}^2\) + \tK_{ij}\tK^{ij} ((d-3)A_{d} -B_{d})^2 \) \bigg] \,.
\nonumber
\eeqa
Noting the absence of $\tK_{ij}\tK^{ij}$ term in eq.~\eqref{eq:CV_full}, we can fix
\begin{equation}
\label{eq:fixB}
B_{d}=(d-3)\, A_{d}\,.
\end{equation}
Further, comparing eqs.~\eqref{eq:CV_full} and \eqref{eq:effectiveWK}, the last parameter is fixed as
\begin{equation}\label{eq:fixA}
A^2_{d}  = \frac{4(d-4)}{(d-2)^2(d-3)}   \,.
\end{equation} 
Finally, we can write the $K$-term \reef{scrabble4} as 
\beqa\label{eq:fixWK}
\widetilde{W}_{K}( \wmB)&=&\frac{4(d-4)}{(d-2)^2(d-3)}  \int_{\wmB} d^{d-1}\sigma \,\sqrt{\tilde{h}}\,\frac{\partial^2 \mathbf{L}_{\rm eff}}{\partial \tR_{ijkl}\partial \tR^{mnop}}\\
&&\qquad\times\ \  \tK_{jl}\(  \tilde{h}_{ik} + (d-3) \tn_i \tn_k\) \tK^{np} \( \tilde{h}^{mo} + (d-3) \tn^m \tn^o \) \,.
\nonumber
\eeqa

Although we have a successful match here, we should point out that the $K$-term defined in eq.~\eqref{eq:defineWK} was chosen for its simplicity and in a similar spirit to the analogous term appearing in the Wald-Dong entropy. However, it is easy to find many other ways in contracting all the indexes in $\frac{\partial^2 \mathbf{L}_{\rm eff}}{\partial \tR_{ijkl}\partial \tR^{mnop}}$ with two extrinsic curvatures and combinations of $\tdh^{ij}$ and $\tn^i \tn^j$. Some examples would include
\begin{equation}
\begin{split}
  &\frac{\partial^2 \mathbf{L}_{\rm eff}}{\partial \tR_{ijkl}\partial \tR^{mnop}} \tK_{ik} \tK_{jl}\(  A_1 \,\tg^{mo} +  B_1 \,\tn^m \tn^o\) \( A_1 \,\tg^{np} + B_1\, \tn^n \tn^p \) \,,\\
  &\frac{\partial^2 \mathbf{L}_{\rm eff}}{\partial \tR_{ijkl}\partial \tR^{mnop}} \tK_{i}^m \tK_{j}^n\(  A_2\, \tg_{kl} +  B_2 \,\tn_k \tn_l\) \( A_2 \,\tg^{op} + B_2\, \tn^o \tn^p \) \,.\\
\end{split}
\end{equation}
Note that in the first case, both extrinsic curvatures are contracted with the indices of a single variation with respect to the Riemann tensor, while in the second, the two indices of each individual extrinsic curvature are contracted with different variations. Note that no terms with these structures appear in the $K$ corrections of the Wald-Dong entropy \cite{Dong:2013qoa}. However,
at present, we do not have a strong reason to rule out these expressions or their linear combinations. This means that in general, there is much more ambiguity in defining $\widetilde W_K(\wmB)$ than indicated in eq.~\reef{scrabble4} and the numerical coefficients can not be completely fixed. This stands in contrast with the Wald-Dong entropy, for which a unique extrinsic curvature term is derived from the replica trick \cite{Dong:2013qoa}. Unfortunately, we do not have a proper derivation of the complexity=volume proposal, which we might extend to probe the complexity of theories dual to higher derivative gravity.  However, we will test our simple ansatz in section~\ref{sec:GB}  by continuing to show that our calculations are consistent with higher curvature gravity in the bulk.

We should also add that we expect that eq.~\reef{eq:fixWK} is only the first in an infinite series of corrections involving the extrinsic curvatures, as appears in the Wald-Dong entropy. Here, we have limited ourselves to the terms quadratic in $\tilde K$ because we only evaluated the effective action \reef{eq:effectiveaction} to include the terms which are quadratic in the curvatures. It may be interesting to extend our calculations to third order, from which we expect to find $\tilde K^3$ contributions to $\widetilde W_K$.

In summary, we find that the leading contributions from the geometry in the vicinity of the brane from usual subregion-CV proposal for the bulk Einstein gravity suggests a generalized CV formula for the induced gravity theory on the brane, \ie 
\begin{equation}\label{eq:CVWWK}
\text{ext}\[ \frac{V( \mathcal{B})}{\Gbulk\, \ell}    \] \simeq  \frac{\widetilde{W}_{\rm gen}\big( \wmB \big) + \widetilde{W}_K (\wmB)}{\Geff\,\ell'}   \,,
\end{equation}
where the generalized volume $\widetilde{W}_{\rm gen}$ and $\widetilde{W}_K$ term are fixed in eqs.~\eqref{eq:fixW} and \eqref{eq:fixWK}, respectively. Further, the scales, $\ell$ in the bulk and $\ell'$ on the brane, are related by eq.~\reef{newel}. We should stress that the above identification relies on the extremality of the bulk surface ${\cal B}$, which was required in deriving eq.~\reef{eq:full_subCV}.
As commented above, we propose that these results can be used to generalize the holographic complexity=volume conjecture for higher curvature gravity theories in general, as in eqs.~\reef{eq:ourporposal} and \reef{eq:details}. Further,  we will test this proposal in section~\ref{sec:GB},  by examining our holographic model with higher curvature gravity in the bulk.

\subsubsection{DGP term on the brane}\label{subsec:DGP}

In a construction analogous to that of Dvali, Gabadadze and Porrati (DGP)~\cite{Dvali:2000hr}, one can also add an intrinsic Einstein term to brane action as follows -- for details see \cite{Chen:2020uac}
\begin{equation}\label{eq:DGP}
I_{\rm brane}= -(T_o-\Delta T) \int d^d x \sqrt{-\tg} + \frac{1}{16\pi G_{\rm{brane}}} \int d^d x \sqrt{-\tg} \, \tR \,,
\end{equation}
which yields the new effective gravitational action on $d$-dimensional brane as
\begin{equation}\label{eq:effectiveaction02}
\begin{aligned}
I_{\rm {eff}}=\frac{1}{16 \pi G_{\rm{eff}}} & \int d^{d} x \sqrt{-\tilde{g}}\left[\frac{(d-1)(d-2)}{\ell_{\text {eff }}^{2}}+\tilde{R}(\tilde{g})\right] \\
&+\frac{1}{16 \pi G_\mt{RS}} \int d^{d} x \sqrt{-\tilde{g}}\left[\frac{L^{2}}{(d-4)(d-2)}\left(\tilde{R}^{i j} \tilde{R}_{i j}-\frac{d}{4(d-1)} \tilde{R}^{2}\right)+\cdots\right] \,.
\end{aligned}
\end{equation}
In the first line of this action, the new effective Newton constant associated with Einstein term is given by 
\begin{equation}\label{scrabble1}
\frac{1}{G_{\rm{eff}}}= \frac{2L}{(d-2)\Gbulk} + \frac{1}{\Gbrane}  \,,
\end{equation}
while in the second line, $G_\mt{RS}=(d-2)\Gbulk/(2L)$.

This provides an interesting framework to extend our generalized proposal for complexity=volume. In the case of holographic entanglement entropy, one can clearly argue that the DGP term introduces a brane contribution in eq.~\reef{eq:island2} by simply following the derivations in \cite{Myers:2010tj,Lewkowycz:2013nqa}.
Unfortunately, such a derivation is lacking for the CV formula, and so we will simply say that it is natural to expect that with a DGP term, the CV proposal should have a similar extension to include a contribution proportional to the volume of $\wmB=\mB\,\cap\,$brane. More precisely, if the extremal surface crosses a DGP brane, then eq.~\reef{eq:CV-notboth} would become
\begin{equation}
\label{eq:CV-both}
{\cal C}_{\mt{V}}^{\rm{sub}}(\mA) = \max_{\partial {\cal B}=\mA\cup {\RT}}   \[\frac{V(\mB)}{\Gbulk\, \ell} + \frac{V(\wmB)}{\Gbrane \,\ell'}\]\,,
\end{equation}
where $\ell$ and $\ell'$ are the independent `unknown' length scales for the bulk and brane, as are expected for the CV ansatz.

Now if we examine the leading contributions from the bulk geometry in the vicinity of the brane, as in eq.~\reef{scrabble0}, the above expression yields
\begin{equation}\label{eq:braneCV}
\mC^{\rm{sub}}_\mt{V} \( \mA\)  = \frac{2L\, V(\wmB) }{(d-1)\Gbulk \ell} +\cdots + \frac{V(\wmB)}{\Gbrane \,\ell'} =\frac{ V(\wmB)}{\Geff\, \ell'} + \cdots 
\end{equation}
where to produce the second equality, we have used eq.~\reef{newel} to relate the two length scales, $\ell$ and $\ell'$, and then eq.~\reef{scrabble1} applied for effective Newton's on the brane. Hence, we see that combining eqs.~\reef{newel} and \reef{eq:CV-both} produces a consistent framework with which to understand complexity=volume for the brane theory. While we have ignored the higher curvature terms above, it is clear that including the DGP term on the brane leads to the same results as eqs.~\eqref{eq:fixW} and \eqref{eq:fixWK} with the same dimensionless coefficients for the new gravity theory \reef{eq:effectiveaction02} on the brane. It would be interesting to examine if this approach continues to succeed if one were to extend the brane action \reef{eq:DGP} with higher curvature terms.

\subsection{Maximal islands}\label{subsec:maximal}

Up to this point, we have shown that with the usual subregion-CV proposal \reef{eq:CV-notboth}\footnote{Or using eq.~\reef{eq:braneCV} for the case of a DGP brane.} and applying the FG expansion for extremal surfaces in the bulk, integrating the leading contributions in the vicinity of the brane produces a  generalized CV formula for the induced theory on the brane. 
In particular, the new complexity functional \reef{eq:new} is easily derived from the higher-curvature gravity action on the brane \reef{eq:effectiveaction} using eqs.~\eqref{eq:fixW} and \eqref{eq:fixWK}. We stress that the above identification relies on the extremality of the surface ${\cal B}$ in the bulk, which was required in deriving eq.~\reef{eq:full_subCV}. However, at this point, we want to turn to the appearance of the maximization that appears in eq.~\reef{eq:new}.

Here it is enlightening to return to the relation between the island rule \reef{eq:islandformula} on the brane and the RT prescription \reef{eq:island2} in the bulk -- see discussions in \cite{Chen:2020uac,Chen:2020hmv}. Our first observation is that analogous to our analysis above, carefully examining the extremal RT surfaces near the brane shows that the Bekenstein-Hawking formula in the island rule \eqref{eq:islandformula} actually expands to the Wald-Dong entropy for the gravity action induced on the brane~\cite{Chen:2020uac}. As in the above, this requires that we solve the local equations in the bulk which extremize the RT surfaces away from the brane, but in doing so, one produces a family of solutions that are extremal in the bulk (and have the fixed boundary conditions on the asymptotic AdS boundary) but which have different profiles on the brane. Finding the correct solution amongst this family can be characterized in terms of satisfying a particular boundary condition at the brane -- see eq.~(4.17) in \cite{Chen:2020uac}. However, a more pragmatic approach is to simply find the correct solution by varying over the possible profiles on the brane to see which one actually minimizes the entropy functional in eq.~\reef{eq:island2}. This second stage is then precisely the extremization appearing in the island rule \reef{eq:islandformula}.

Of course, the same narrative applies here to the holographic complexity. Recall that our boundary state was defined on a region $\mA=\mA_\mt{L}\cup\mA_\mt{R}$, where the subregions $\mA_\mt{L,R}$ sit to either side of the conformal defect in the asymptotic boundary, as shown in figure \ref{twoAdS}. Similarly, we divide the bulk surface ${\cal B} ={\cal B}_\mt{L}  \cup {\cal B}_\mt{R}$ into the two components on either side of the brane. For both of these components, we demand that these surfaces are extremal away from the brane by solving eq.~\reef{eq:extremalEinstein}, subject to the boundary condition that ${\cal B}_\mt{L,R}$ are anchored at the corresponding  $\mA_{\mt{L, R}}$ on the asymptotic boundary, the RT surface $\Sigma_{\mA}$ in the bulk, and the island $\wmB$ on the brane, 
\ie $\partial {\cal B}_\mt{L}=\mA_\mt{L}\cup \Sigma_\mA\cup \wmB$ (and similarly for the right side). In particular, both surfaces $\mB_\mt{L,R}$ intersect the brane along with the common profile $\wmB$, however, this profile is left undetermined at this stage. Hence we find a wide family of codimension-one surfaces which are extremal in the bulk, \ie away from the brane. Then, to find to correct extremal surface, we must finally maximize that volume by varying over the possible profiles. That is, we have decomposed the extremization of $\mB$ into two steps: 
\begin{equation}\label{revolver}
\mC_{\mt{V}}^{\rm sub} (  \mA ) \equiv \max_{\partial\wmB=\RTbrn}\! \( \underset{\mB_{\mt{L}},\mB_{\mt{R}}}{\text{ext}} \[ \frac{ V( \mathcal{B}_{\mt{L}})+V( \mathcal{B}_{\mt{R}}) }{\Gbulk \ell}   \] \)\,.
\end{equation}
Combined with the near-brane contributions in eq.~\eqref{eq:CVWWK}, this equation then becomes
\begin{equation}\label{eq:CVequalsisland}
\mC_{\mt{V}}^{\rm sub} (  \mA)  = \max_{\partial\wmB=\RTbrn} \[ \frac{\widetilde{W}_{\rm gen}( \wmB) + \widetilde{W}_K (\wmB) }{G_{\rm eff}\ell'}  +\cdots \]  \,,
\end{equation}
using the generalized volume and $K$-term in eqs.~\eqref{eq:fixW}  and \eqref{eq:fixWK}, respectively. 

The ellipsis in eq.~\reef{eq:CVequalsisland} indicates the contributions coming far from the brane, \ie from regions with $\theta_\mt{B}\ll \theta\le \pi$ with the coordinates in eq.~\reef{metric33}. It is interesting to note that  the analogous contributions for the holographic entanglement entropy \reef{eq:island2} provide the quantum contributions when interpreted in terms of the effective $d$-dimensional brane perspective, \ie $S_\mt{QFT}( \bdyReg \cup \islands )$ in eq.~\reef{eq:islandformula}. Hence it is natural to expect that the corresponding contribution in the holographic complexity constitutes a (semiclassical) contribution in the bath region R combined with the island $\wmB$ on the brane. We return to discuss this point in section \ref{sec:dis}.


%% file: sections/Gauss.tex

In the previous section, we showed how holographic complexity naturally arises for the induced gravity theory on the brane in the doubly holographic model of \cite{Chen:2020uac,Chen:2020hmv}. 
However, beginning with the usual complexity=volume conjecture \reef{eq:CVsubregion} for  ordinary Einstein gravity in the bulk, we were lead to a generalization of the CV proposal suitable for higher curvature gravity, such as the induced theory \reef{eq:effectiveaction} on the brane. Our proposal is that the new functional appearing for the holographic complexity on the brane should in fact serve to provide a generalized complexity=volume conjecture for any higher curvature theory
\begin{equation}\label{eq:defCV}
\mC_\mt{V}^{\rm sub} (\mA) = \max_{\partial \mB={\mA}\cup {\RT}} \[  \frac{W_{\rm gen} ( \mB )+ W_K (\mB)}{G_{\mt{N}}\,\ell}  \]  \,,
\end{equation}
with the functionals given in eq.~\reef{eq:details}.
As indicated, the maximization is performed over all possible codimension-one surfaces $\mB$ anchored at the subregion $\mA$ on the asymptotic boundary and the corresponding RT surface $\Sigma_{\mA}$ in the bulk. Of course, this proposal reduces to the standard CV conjecture~\eqref{eq:CVsubregion} when the bulk theory is Einstein gravity. 

In this section, we examine a new consistency check for our new proposal by considering higher curvature gravity in the bulk. That is, we start by considering a theory of higher curvature gravity in the ($d+1$)-dimensional bulk and apply eq.~\reef{eq:defCV} for the holographic complexity. Then following the analogous calculations as in section \ref{sec:bulksubregion}, we show that the holographic complexity for the induced theory on the $d$-dimensional brane takes the same form, \ie
\begin{equation}\label{eq:CVequalsisland02}
 \mC_{\mt{V}}^{\rm sub} (  \mA)\simeq 
 \mC_\mt{V}^{\rm Island}= \max_{\partial\wmB=\RTbrn} \[ \frac{\widetilde{W}_\mt{gen}\big( \wmB \big) + \widetilde{W}_K (\wmB)}{\Geff\,\ell'}  \]  \,,
\end{equation}
where the functionals $\widetilde{W}_\mt{gen}$ and $\widetilde{W}_K$ are adapted to the new spacetime dimension and the induced gravity action on the brane.

\begin{figure}[tb]
	\centering\includegraphics[width=4.5in]{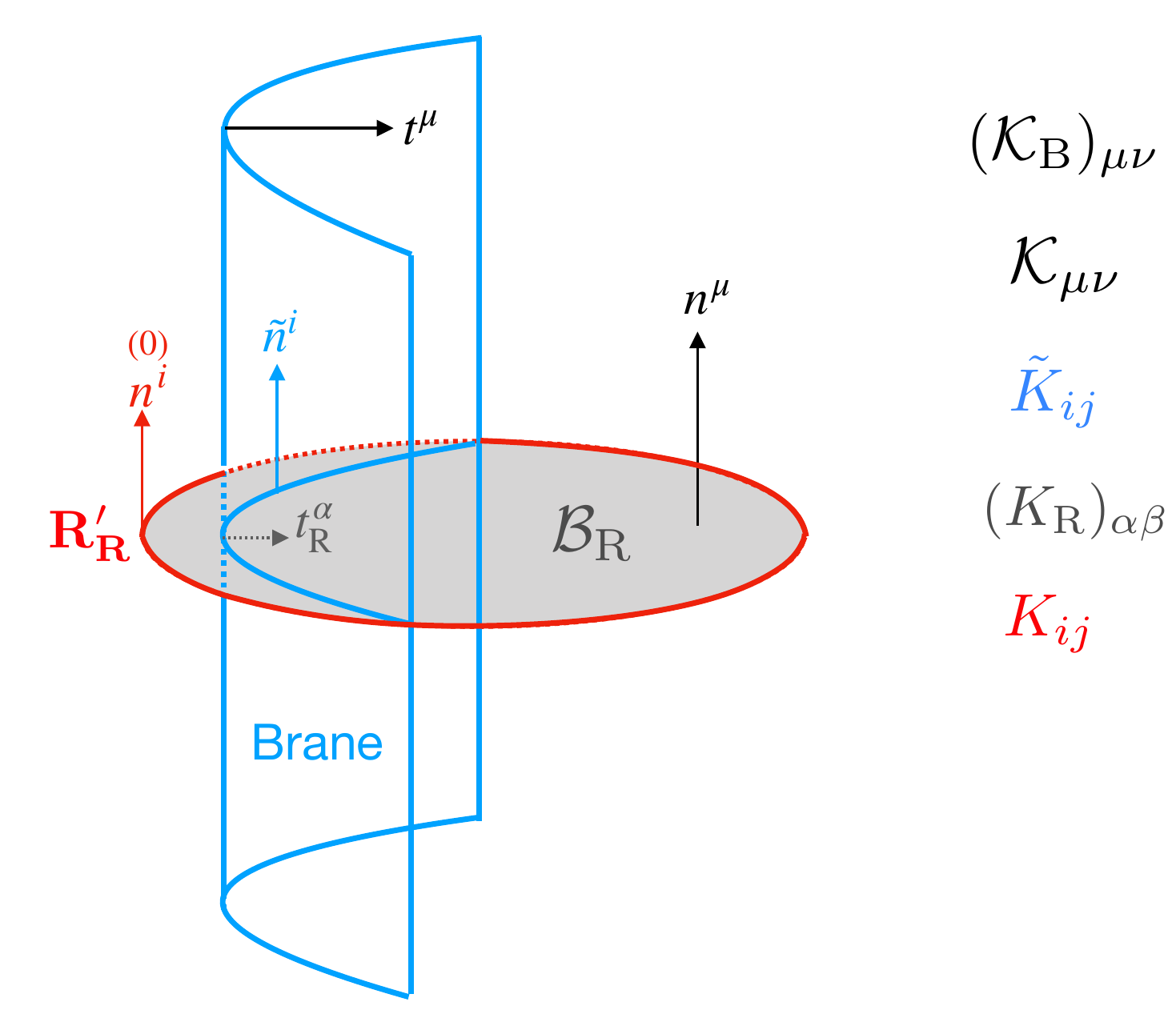}
	\caption{Different hypersurfaces in the doubly holographic system and their corresponding extrinsic curvatures.\label{hypersurface}}
\end{figure}

Our calculations will refer to several different hypersurfaces and the corresponding extrinsic and intrinsic curvatures associated with these surfaces -- see figure \ref{hypersurface}. In order to clarify the notation, we list the different curvatures here:\footnote{One may keep in mind that generally, we use $\mK$ to denote the extrinsic curvature of a $d$-dimensional hypersurface embedded in $(d+1)$-dimensional bulk, while $K$ designates the extrinsic curvature for a $(d-1)$-dimensional hypersurface embedded in a $d$-dimensional submanifold. We also adopt a similar notation for the intrinsic curvatures}
\begin{itemize}
\item  the $(d+1)$-dimensional bulk, with intrinsic curvature $\mR[g^\mt{bulk}_{\mu\nu}]$;
	\item  the spacelike surfaces $\mathcal{B}$ embedded in the $(d+1)$-dimensional bulk, with timelike normal $n^\mu$, extrinsic curvature $\mK_{\mu\nu}$ and intrinsic curvature $R_{\mB}[h_{\alpha\beta}]$;
	\item  the brane embedded in the $(d+1)$-dimensional bulk, with spacelike normal $t^\mu$, extrinsic curvature $(\mK_\mt{B})_{\mu\nu}$ and intrinsic curvature $\tR[\tilde{g}_{ij}]$;
	\item  the island region $\wmB=\mB\,\cap\,$brane (with $\mB=\mB_\mt{L}\cup\mB_\mt{R}$) thought of as being embedded in the surface $\mB_{\mt{R}}$, with spacelike normal $t_{\mt{R}}^\alpha$ and extrinsic curvature $(K_{\mt{R}})_{\alpha\beta}$; similarly for $\wmB$ embedded in the surface $\mB_{\mt{L}}$, we have the spacelike normal $t_{\mt{L}}^\alpha$ to the island and extrinsic curvature $(K_{\mt{L}})_{\alpha\beta}$;\footnote{Note that in general, we will consider surfaces $\mB$ which are not smooth where they cross the brane, \eg before extremizing the profile of $\wmB$ on the brane -- see discussion in section \ref{subsec:maximal}. Hence we must consider embedding $\wmB$ in $\mB_\mt{R}$ and $\mB_\mt{L}$ separately.} 
	\item  the island region $\wmB$ thought of as being embedded in the brane, with timelike normal $\tilde{n}^i$ and extrinsic curvature $\tilde{K}_{ij}$;
	\item  the subregion $\mA'$ (where $\mB_\mt{L,R}$ would meet a virtual asymptotic boundary at $z=0$)\footnote{Recall that the brane cuts off the geometry at some $z_\mt{B}\ll L$, but to employ the FG expansion, we consider extending both the left and right geometries beyond the brane to a virtual asymptotic boundary at $z=0$ -- see discussion in section~\ref{subsec:CVEinstein}. In principle, all of these quantities should also carry a subscript L or R to indicate quantities associated with the geometry on the left or right of the brane.} embedded in the asymptotic boundary, with timelike normal $\ov{0}{n}{}^i$, extrinsic curvature $K_{ij}$ and intrinsic curvature $R_{\Sigma}[\ov{0}{h_{ab}}]$;
\item the virtual asymptotic boundary (see above) with intrinsic curvature $R[\ov{0}{g_{ij}}]$.
\end{itemize}

\subsection{Holographic complexity for Gauss-Bonnet gravity}

Our first consistency check with higher curvature gravity consists of having Gauss-Bonnet gravity in the bulk. The bulk gravitation action is therefore given by
\begin{equation}\label{eq:actionGB}
I_{\rm {bulk}}^{\mt{GB}}=\frac{1}{16 \pi G_{\rm{bulk}}} \int d^{d+1} y\, \sqrt{-g}\left[\frac{d(d-1)}{L^{2}}+\mathcal{R}[g_{\mu\nu}]+\lambda_{\mt {GB}}\, \mL_{\mt{GB}}\right]+I_{\rm{surf}}^{\mt{GB}} \,, 
\end{equation}
with the Gauss-Bonnet term defined by 
\begin{equation}
\lambda_{\mt {GB}}=\frac{L^{2} \lambda}{(d-2)(d-3)}\,,\qquad \mL_{\mt{GB}} = \mR_{\mu\nu\rho\sigma}\mR^{\mu\nu\rho\sigma}-4\mR_{\mu\nu}\mR^{\mu\nu}+\mR^2\,.
\end{equation}
Here, we have explicitly included the boundary term $I_{\text {surf }}^{\mt{GB}}$ to emphasize that GB gravity has a well-posed variational principle with Dirichlet boundary conditions $\delta g_{\mu\nu} =0$ \cite{Lovelock:1971yv}.
Similar to the standard Gibbons-Hawking-York term, the extended boundary term is given by \cite{Myers:1987yn}
\begin{equation}\label{eq:actionGBbdy}
\begin{split}
I_{\text {surf }}^{\mt{GB}}=\frac{1}{16 \pi G_{\rm{bulk}}} \oint d^{d} x \sqrt{-\tilde{g}} &\left[2 \mK_{\mt B}+\frac{4L^{2} \lambda}{(d-2)(d-3)}\left( \tR \mK_{\mt B}-2 \tR_{i j} \mK_{\mt B}^{i j} +  J\right)\right]\,,
\end{split}
\end{equation}
where $\tR_{ij}$ and $\mK_{ij}$ denote the Ricci tensor and extrinsic curvature associated with the boundary geometry, and $J $ is the trace of 
\begin{equation}
J_{ij} \equiv\frac{1}{3} \(  2 \mK\, \mK_{ik}\, \mK^{k}{}_j  + \mK^{kl}\,\mK_{kl} \,\mK_{ij} - 2 \mK_{ik}\,\mK^{kl}\,\mK_{lj} - \mK^2 \mK_{ij}       \) \,.
\end{equation}

The presence of the Gauss-Bonnet term modifies the Israel junction conditions  \eqref{eq:junction} determining the position of the brane as \cite{Davis:2002gn,Deruelle:2000ge}
\begin{equation}\label{eq:JunctionGB}
\Delta (\mK_{\mt{B}})_{ij} - \tilde{g}_{ij}\, \Delta \mK_{\mt{B}} + 2\lambda_{\mt{GB}} \,\Delta\! \[\tilde{E}_{iklj} \mK_{\mt{B}}^{kl} + 3J_{ij}(\mK_\mt{B}) - J\, \tg_{ij}\] =8 \pi \Gbulk\, S_{ij} \,, 
\end{equation}
where the tensor $\tilde{E}^{ijkl}$ is defined as 
\begin{equation}\label{defE}
\tilde{E}^{ijkl}= 2\tR\, \tg^{i[k}\tg^{l]j} - 4 \(  \tR^{i[k}\tg^{l]j} + \tR^{j[l}\tg^{k]i} \)+ 2 \tR^{ijkl}\,.
\end{equation}
This generalized Israel junction condition can be derived by considering a thin shell and taking the thickness of the shell $\delta z \to 0$ -- see \cite{Deruelle:2000ge} for details.  Similar to the derivation of the Israel junction condition for Einstein gravity, one can also obtain the generalized Israel junction condition by considering the gravitational action on either side of the brane with the boundary term in eq.~\reef{eq:actionGBbdy} at the brane \cite{Davis:2002gn}. That is, with these boundary terms, we solve the gravity equations in the bulk away from the brane with some fixed boundary condition for $g_{\mu\nu}$ at the brane (as well as asymptotic infinity, of course). Then we solve the full system by allowing $g_{\mu\nu}$ at the brane surface to vary and gluing the two surfaces together while demanding that the generalized Israel boundary condition in eq.~\reef{eq:JunctionGB} is satisfied. We should note that the latter approach is implicitly adopted in deriving the induced gravity action in eq.~\reef{eq:actionGBbrane} -- see appendix A of \cite{Myers:2013lva}. More specifically, in evaluating the bulk action in the vicinity of the brane, it is essential to include the contribution of the boundary term.\footnote{The same approach was applied in deriving eq.~\reef{eq:effectiveaction} with Einstein gravity in the bulk \cite{Chen:2020uac}.}

While the length scale $L$ defines the cosmological constant in the action \reef{eq:actionGB}, the curvature scale $\tL$ of the AdS vacuum solution in the Gauss-Bonnet gravity is\footnote{We only consider the solution here which yields $\tL\to L$ in the limit $\lambda\to0$.}
\begin{equation}
\tL^2 = \frac{L^2}{f_{\infty}}\,, \qquad \text{with} \qquad f_{\infty} = \frac{1-\sqrt{1-4\lambda}}{2\lambda} \,. 
\end{equation}
The induced gravitational action on the brane is given by \cite{Myers:2013lva} 
\beqa\label{eq:actionGBbrane}
I_{\rm {eff}}^{\mt{GB}}&=&\frac{1}{16 \pi \Geff}\int_{\rm brane} \!\!\!\!d^{d} x\, \sqrt{-\tilde{g}}\,\bigg[\frac{(d-1)(d-2)}{\ell_{\text {eff }}^{2}}+\tilde{R}[\tilde{g}]\\
&&\qquad\qquad\qquad+\kappa_{1}\left(\tR_{i j} \tR^{i j}-\frac{d}{4(d-1)} \tR^{2}\right)+\kappa_{2} \tilde{C}_{i j k l} \tilde{C}^{i j k l}+\cdots\bigg] \,,
\nonumber
\eeqa
where the effective Newton constant and coupling constants are 
\beqa\label{hunter}
&&\qquad\qquad\qquad\quad
 \frac{1}{\Geff}= \frac{2\tL}{ d-2} \frac{1+2\lambda f_\infty}{\Gbulk}\,, \label{conGB}\\
\kappa_1&=&\frac{\tL^2 }{(d-2) (d-4)}\,
\frac{1-6\lambda f_\infty}{1+2\lambda f_\infty} \,, 
\qquad 
\kappa_2= \frac{\tL^2 }{(d-3) (d-4)}\,
\frac{\lambda f_\infty}{1+2\lambda f_\infty}\,,
\nonumber
\eeqa
and $\tilde{C}_{i j k l}$ denotes the Weyl tensor on the brane. We also note that the expression for the scale $\ell_\mt{eff}$ in eq.~\reef{eq:Gd} is replaced by
\begin{equation}
\frac{1}{\ell_\mathrm{eff}^2}= \frac{2}{\tilde{L}^2(1+2\lambda f_{\infty})} \(  1- \frac{2}{3}\lambda f_{\infty} - \frac{4\pi \tilde{L}\Gbulk T_o }{d-1} \) \,.
\end{equation}
In the following, we adopt our proposal  \eqref{eq:defCV} to evaluate the holographic complexity for $(d+1)$-dimensional GB gravity in the bulk and compare the leading terms in the FG expansion near the brane to the complexity of the island in the $d$-dimensional effective higher-curvature gravity on the brane. As we will see below, the leading terms in the generalized holographic CV for the boundary subregion agree with the proposed complexity \reef{eq:CVequalsisland02} of the island. We see this consistency as extra support for our proposal for holographic complexity for higher-curvature gravity theory. 

In evaluating the holographic complexity for ($d$+1)-dimensional bulk theory, we consider a codimension-one slice $\mathcal{B}$, with time-like normal $n^\mu$ and induced metric $h_{\mu\nu} = g_{\mu\nu} + n_{\mu}n_{\nu}$. Following the analysis in the previous section, our first step is to extremize the complexity functional on $\mB$ away from the brane, while leaving the profile $\wmB$ on the brane undetermined. In order to ensure that this involves a well-defined variational principle, we actually extend eq.~\eqref{eq:defCV} to include a surface term
\begin{equation}\label{eq:CV_GBbulk}
	\mC_\mt{V}^{\rm sub} (\mA)= \max_{\partial \mB={\mA}\cup {\RT}} \[  \frac{ W_{\rm gen}(\mathcal{B}) + W_K(\mathcal{B})   + W_{\rm bdy}(\partial\mB_\mt{L}\cup\partial\mB_\mt{R}) }{G_{\rm{bulk}}\ell}  \]   \,.
	\end{equation}
Of course, the generalized volume $ W_{\rm gen}$ and the $K$-term are defined in eq.~\reef{eq:details}. We do not specify the details of $W_{\rm bdy}$ but its form will become evident in the following. Further, we note that we are evaluating this expression on $\partial\mB_\mt{L}\cup\partial\mB_\mt{R}$. In particular, this contribution appears on (either side of) $\wmB$, which is not really a boundary of the full surface $\mB=\mB_\mt{L}\cup\mB_\mt{R}$. Hence we are treating $W_{\rm bdy}$ in a manner to the gravitational surface \eqref{eq:actionGBbdy}, which appears on either side of the surface defined by the brane -- see discussion below eq.~\reef{defE}. 

Let us begin by evaluating $W_{\rm gen}$ for the GB theory. It is straightforward to obtain
\begin{equation}\label{eq:GBWgen}
\begin{split}
&\quad  \alpha_{d+1} \frac{\partial (\mathcal{R}+\lambda_{\mt {GB}} \mL_{\mt{GB}})}{\partial \mR_{\mu\nu\rho\sigma}} n_{\mu} h_{\nu\rho} n_{\sigma}+\gamma_{d+1} \\
&=  \frac{2 (d-3)}{(d-1)(d-2)} \( \frac{d}{2}  + \lambda_{\mt{GB}} (d-2)( \mR + 2\mR^{\mu\nu} n_{\mu}n_{\nu}) \)+\frac{2 }{(d-1)(d-2)} \\
&=  1+ \lambda_{\mt{GB}}\frac{2(d-3)}{(d-1)}\( \mR +2 \mR^{\mu\nu} n_{\mu}n_{\nu}\)  \,,
\end{split}
\end{equation}
where the values of $\alpha_{d+1}, \gamma_{d+1}$ are given using eq.~\eqref{eq:alphagamma}. Now using eq.~\eqref{eq:secondderivative}, the $W_K$ term yields
\begin{equation}\label{eq:GBWK}
\begin{split}
&  A_{d+1}^2\(\frac{\partial^2  (\mathcal{R}+\lambda_{\mt {GB}} \mL_{\mt{GB}})}{\partial \mR_{\mu_1\nu_1\rho_1\sigma_1}\partial \mR_{\mu_2\nu_2\rho_2\sigma_2}} \mK_{\nu_1\sigma_1}\(  h_{\mu_1\rho_1} + (d-2) n_{\mu_1} n_{\rho_1}\) \mK_{\mu_2\sigma_2} \(h_{\mu_2\sigma_2} + (d-2) n_{\mu_2} n_{\sigma_2}\) \)\,,\\
&= \frac{4\lambda_{\mt{GB}}(d-3) }{(d-1)^2(d-2)} \(  \frac{\mK^2}{2}  - \frac{\mK^2}{2} \( (d+1)(d-4) +8\)  +\frac{1}{2}\( \mK^2 +(d-1)(d-2) \mK_{\mu\nu}\mK^{\mu\nu} \) \) \,,\\
&= \frac{2 \lambda_{\mt{GB}}(d-3)}{(d-1)}\(  \mK^{\mu\nu}\mK_{\mu\nu} - \mK^2 \) \,,
\end{split}
\end{equation}
where $A_{d+1}^2$ was replaced using eq.~\eqref{eq:fixA}.
Noting Gauss's ``Theorema Egregium" for the hypersurface $\mB$ with the induced metric $h_{\alpha\beta}$ and intrinsic curvature $R_{\tiny {\mB}}$, \ie 
\begin{equation}
 R_{\mB}[h_{\alpha\beta}] = \mR[g_{\mu\nu}] + \( 2\mR^{\mu\nu}n_{\mu}n_{\nu} -\mK^2 + \mK_{\mu\nu} \mK^{\mu\nu}\) \,,
 	\end{equation}
 we can recast the $\lambda_{\mt{GB}}$-terms into the intrinsic geometric quantities of hypersurface $\mB$, \ie
 \begin{equation}\label{eq:WgenWK}
 W_{\rm gen}(\mathcal{B})  + W_{K}(\mathcal{B})  = \int_{\mB} d^{d-1}\sigma dz\sqrt{\det h_{\alpha\beta}}  \(  1+ \frac{2L^2 \lambda }{(d-1)(d-2)} R_{\mB} \) \,. 
 \end{equation}
 
Given the above result, it is straightforward to derive the desired $W_{\rm bdy}$.\footnote{It would also be natural to derive $W_{\rm bdy}$ from the surface terms $I_{\text {surf }}^{\mt{GB}}$ in the gravitational action appearing on either side of the brane. However, our initial attempts were unsuccessful and indicate that there are subtleties in this approach.} Namely, extremizing this generalized volume functional will have a good variational principle if we add the usual ‘Gibbons-Hawking’ term on the boundary.  The island contribution on the brane  is then given by 
 \begin{equation}\label{eq:GB_Wbdy}
 W_{\rm bdy}  ( {\wmB})= \frac{4L^2 \lambda }{(d-1)(d-2)}  \int_{{\wmB}} d^{d-1}\sigma \sqrt{\det \tilde{h}} \,  (K_{\mt{L}}+K_{\mt{R}}) \,,
 \end{equation}
where $K_{\mt{L}}, K_{\mt{R}}$ denote the trace of the extrinsic curvature of $\wmB$ embedded in $\mB_{\mt{L}}, \mB_{\mt{R}}$, respectively.\footnote{We may note the importance of this term by observing that the GB contribution in eq.~\eqref{eq:WgenWK} is negative for $\lambda>0$ because $\RB$ is negative. On the other hand, in eq.~\reef{hunter}, we see the effective Newton constant on the brane has a positive contribution from the Gauss-Bonnet term. Hence for the corrections of the GB term to the coefficient of the volume term in the holographic complexity on the brane to match, there must be an additional contribution beyond eq.~\eqref{eq:WgenWK}. Indeed, we will find the extra contribution from eq.~\eqref{eq:GB_Wbdy} yields the desired match.} Similar to the extremizing condition for entanglement entropy in GB gravity (\eg see \cite{Chen:2013qma,Bhattacharyya:2014yga}), it is straightforward to find that the generalized CV functional \reef{eq:CV_GBbulk} for GB gravity is extremized by the following local equation  
\begin{equation}\label{eq:extremalGB}
 \mK   +  \frac{2L^{2} \lambda}{(d-2)(d-3)} \(  R_{\mB} \, \mK - 2R_{\mB}^{\alpha\beta} \mK_{\alpha\beta} \)    =  0 \,.
\end{equation}
This extremizing condition generalizes eq.~\eqref{eq:WgenWK} for Einstein gravity to GB gravity in the bulk and ensures that $\mB$ is extremal away from the brane.

In summary, applying our proposal \eqref{eq:CV_GBbulk} to GB gravity in the bulk, the generalized holographic complexity becomes
\begin{equation}\label{eq:CVGBbulk}
\begin{split}
\mC_\mt{V}^{\rm sub} (\mA)&= \max_{\partial \mB={\mA}\cup {\RT}} 	\frac{1}{G_{\rm{bulk}}\ell}\bigg[\,V(\mathcal{B})  \\
& \left.   +  \frac{2L^2 \lambda }{(d-1)(d-2)}  \( \int_{\mB} d^{d-1}\sigma dz\sqrt{\det h} \, R_{\mB} +  2\int_{\wmB} d^{d-1}\sigma \sqrt{\det \tilde{h}} \, ( K_{\mt{L}} +K_{\mt{R}}) \)  \right]\,, 
\end{split}
\end{equation}
and the resulting condition for extremality of $\cal B$ in the bulk is given by eq.~\eqref{eq:extremalGB}.




\subsubsection{Holographic complexity from induced gravity}
Our goal is to compare the near-brane contributions of  eq.~\eqref{eq:CVGBbulk} to the proposed holographic complexity \eqref{eq:new} on the brane. Hence taking the effective action on the brane in eq.~\eqref{eq:actionGBbrane}, we must evaluate
	\begin{equation}\label{eq:CVGN_brane}
	\mC^{\rm Island}_\mt{V} \equiv  \max_{\partial\wmB=\RTbrn} \[\frac{\widetilde{W}_{\rm gen}\big(\wmB\big) + \widetilde{W}_{K}\big(\wmB\big)  }{\Geff\,\ell'}   \]\,,
	\end{equation}
where the generalized volume and $K$-term are defined in eqs.~\eqref{eq:fixW} and \eqref{eq:fixWK}. The boundary term $\widetilde{W}_{\rm bdy}$ does not affect the calculation of the complexity of the island. In fact, most of $\mC_\mt{V}^{\rm Island}$ is the same as that found in section~\ref{sec:bulksubregion} (see eq.~\eqref{eq:CV_full}) except for the contributions from $\tilde{C}_{i j k l} \tilde{C}^{i j k l}$ term in \eqref{eq:actionGBbrane}. Noting the square of Weyl tensor reads 
\begin{equation}
\tilde{C}_{ijkl}\tilde{C}^{ijkl} = \tR_{ijkl} \tR^{ijkl} -\frac{4}{d-2} \tilde{R}_{ij} \tilde{R}^{ij} + \frac{2}{(d-1)(d-2)} \tR^2\,,
\end{equation}
and using eq.~\eqref{eq:firstderivative} again, the following tensor contraction gives
\begin{equation}\label{eq:zeroWeyl}
\begin{split}
&\quad \frac{\partial ( \tilde{C}_{i j k l} \tilde{C}^{i j k l})}{\partial \tR_{ijkl}} \tn_{i} \tilde{h}_{jk} \tn_{l} \\
&= \(  -2\tR^{ab}\tn_a\tn_b -\frac{4}{(d-2)} \frac{1}{2}\( \tR -\tR^{ab}\tn_a\tn_b(d-2) \)+ \frac{2}{(d-1)(d-2) }\tR \)\\
&=0\,,
\end{split}
\end{equation}
where we show the individual contributions from $\tR_{ijkl} \tR^{ijkl} , \tR_{ij} \tR^{ij}$ and $\tR^2 $, respectively, on the second line. Although the Weyl tensor term does not contribute to the generalized  volume, it still plays a role in the $K$-term. Using eq.~\eqref{eq:secondderivative} and also
\begin{equation}
\frac{\partial^2 (\tR_{i_1j_1k_1l_1}\tR^{i_1j_1k_1l_1})}{\partial \tR_{ijkl} \partial \tR^{mnop}} =2\frac{\partial \tR_{mnop}}{\partial \tR_{ijkl}} \equiv 2(\partial \tR)^{ijkl}_{mnop}   \,,\\
\end{equation}
we find the new contribution to $\tilde{W}_{\tilde{K}}$ from the Weyl-tensor-squared term in the induced action is given by 
\begin{equation}
\begin{split}
 &\kappa_2 A^2_{d} \frac{\partial^2 ( \tilde{C}_{i j k l} \tilde{C}^{i j k l})}{\partial \tR_{i_1j_1k_1l_1}\partial \tR_{i_2j_2k_2l_2}} \tK_{j_1l_1}\(  \tilde{h}_{i_1k_1} + (d-3) \tn_{i_1} \tn_{k_1}\) \tK_{j_2l_2}\(  \tilde{h}_{i_2k_2} + (d-3) \tn_{i_2} \tn_{k_2}\)  \\
 &= \kappa_2 A^2_{d}\[  \frac{1}{2} \(  \tK^2 +(d-2)(d-3)\tK_{ij}\tK^{ij} \) -\frac{4}{(d-2)} \frac{\tK^2}{8}(d(d-5)+8) + \frac{\tK^2}{(d-1)(d-2)} \] \\
 &= \frac{2\lambda f_\infty \tL^2}{(d-2)(d-3)(1+2\lambda f_{\infty})} \(  \tK^{ij}\tK_{ij} -\frac{\tK^2}{d-1} \)\,. 
\end{split}
\end{equation}
Collecting these results, we finally find that the holographic complexity on the island takes the following form 
\begin{equation}\label{eq:CVGBbrane}
\begin{split}
	\mC_\mt{V}^{\rm Island} &= \max_{\partial\wmB=\RTbrn}\ \frac{1}{G_{\rm eff}\ell'}\left[	V(\wmB) \right.\\
	&+	\frac{\tL^2}2\,\frac{ 1-6\lambda f_\infty}{1+2\lambda f_{\infty}}\int_{\wmB} d^{d-1}\sigma \,\sqrt{\tilde{h}}\(  \frac{\tilde{K}^2}{(d-1)(d-3)}  -\frac{\tilde{R}[\tilde{g}]+ 2\tilde{R}_{ij}[\tilde{g}]\tilde{n}^i\tilde{n}^j}{(d-2)(d-3)} \) \\
   &\left.+\frac{2 \tL^2}{(d-2)(d-3)}\,\frac{\lambda f_\infty}{1+2\lambda f_{\infty}} \int_{\wmB} d^{d-1}\sigma \,\sqrt{\tilde{h}}\(  \tK^{ij}\tK_{ij} -\frac{\tK^2}{d-1} + \mathcal{O}(\zB^4) \) \right]\,. 
\end{split}
\end{equation}
Of course, the above result reproduces the holographic complexity derived from Einstein gravity in the bulk, \ie eq.~\eqref{eq:CV_full}, after setting $ \lambda=0$.



\subsubsection{Holographic complexity from near-brane region}

To compare eq.~\eqref{eq:CVGBbrane} with eq.~\eqref{eq:CVGBbulk}, we must integrate the latter over the bulk region near the brane.  Hence as in the previous section, we turn to the FG expansion and evaluate quantities for $z=z_\mt{B}\ll L$.  

From the FG expansion of the induced metric $h_{\alpha\beta}$ on the time slice $\mB$, \ie 
\begin{equation}
h_{zz}= \frac{\tL^2}{z^2} + \delta h_{zz} \,, \qquad   h_{ab}=\frac{\tL^2}{z^2}\,\ov{0}{h_{ab}} +\delta {h}_{ab}\,,
\end{equation}
one can derive the FG expansion of the Ricci tensor on $\mB$ as
\begin{equation}\label{eq:nearAdS}
 (\RB)^{zz}= -\frac{(d-1)z^2}{\tL^4}+\cdots \,,\quad  (\RB)^{ab}= -\frac{(d-1)z^2}{\tL^4}  \ov{0}{h^{ab}}+\frac{z^4}{\tL^4} (R_{\Sigma})^{ab}\big[\ov{0}{h}\big] +\cdots \,.
\end{equation}
We can see that these curvatures correspond very nearly to those of AdS$_{d}$ with a curvature scale $\tL$. Hence, the extremality condition \eqref{eq:extremalGB} for GB gravity simply reduces to 
\begin{equation}
\(  1- 2 \lambda_{\mt{GB}} \frac{(d-1)(d-2)}{\tL^2} + \mathcal{O}(z^2) \) \mK =0\,. 
\end{equation}
Using the expansion of $\mK$ in eq.~\eqref{eq:exapndEOM}, we find the leading order terms in the FG expansion of the embedding of $\cal B$
\begin{equation}\label{eq:X1GB}
\ov{1}{x^i} \( \sigma^a\)= \frac{\tL^2}{2(d-1)} K n^i  \,,
\end{equation}
which is essentially the same as for Einstein gravity. 
Similarly, we can find the expansion of the induced metric on  $\mB$ 
\begin{equation}
\begin{split}
\delta h_{zz} &= \ov{1}{h_{zz}} + \mathcal{O}(z^2)=- \frac{\tL^2}{(d-1)^2} \, K^2 + \mathcal{O}(z^2) \,, \\
\delta h_{ab}&= \ov{1}{h_{ab}} + \mathcal{O}(z^2)= \ov{1}{g_{ab}} + \frac{\tL^2}{d-1} \, K K_{ab} + \mathcal{O}(z^2)\,.
\end{split}
\end{equation}
To find the subleading contributions in $\RB$, we consider the FG-expansion as a perturbation on the metric $\ov{0}{h_{\alpha\beta}}$ and calculate the perturbation of the Ricci scalar by
\begin{equation}
\delta \RB = - (\RB)^{\alpha\beta} \delta h_{\alpha\beta} + \nabla^\alpha  \nabla ^\beta \delta h_{\alpha\beta} - \nabla ^\alpha \nabla _\alpha \delta h^\beta_\beta  \,.
\end{equation}
Keeping in mind that the terms with more $z$-derivatives dominate in the small $z$ expansion, one can get the expansions of the Christoffel symbols as
\begin{equation}
\Gamma^{z}_{zz} \approx -\frac{1}{z} \, \quad \Gamma^{z}_{ab} \approx \frac{1}{z}\, \ov{0}{h_{ab}}\, \quad \Gamma^{a}_{bz}  \approx   - \frac{1}{z}\,\delta^a_b\,. 
\end{equation}
The Ricci scalar near the asymptotic boundary is given by
\begin{equation}\label{eq:RB02}
\begin{split}
 \RB [h_{\alpha\beta}] &= - \frac{d(d-1)}{\tL^2} + \frac{z^2}{\tL^2} R_{\Sigma} [\ov{0}{h_{ab}}] + \delta \RB  \\
 &=- \frac{d(d-1)}{\tL^2} + \frac{z^2}{\tL^2}  \(  R_{\Sigma}  + \frac{(d-1)(d-2)}{\tL^2} \ov{1}{h_{zz}}  + \frac{2(d-2)}{\tL^2} \ov{0}{h^{ab}} \ov{1}{h_{ab}}   \) + \mathcal{O}(z^4)  \\
 &\approx  -\frac{d(d-1)}{\tL^2} + \frac{z^2}{\tL^2} \[   R_{\Sigma}[\ov{0}{h_{ab}}]  - 2 \ov{0}{h^{ab}} R_{ab} +R+ \frac{(d-2)}{(d-1)} K^2  \] \,,
\end{split}
\end{equation}
with Ricci tensor $R_{ab}$ associated with boundary metric $\ov{0}{g_{ij}}$. We can use the Gauss-Codazzi equation 
\begin{equation}
(R_{\Sigma})_{abcd}= R_{abcd} -  K_{ac}K_{bd} + K_{ad}K_{bc} \,,
\end{equation}
to rewrite the expansion of $\RB [h_{\alpha\beta}]$ as 
\begin{equation}\label{eq:RB}
\RB [h_{\alpha\beta}] \approx  -\frac{d(d-1)}{\tL^2} + \frac{z^2}{\tL^2} \(  K_{ab}K^{ab} -\frac{1}{(d-1)} K^2  \)  + \mathcal{O}(z^4) \,.
\end{equation}
Note that the Ricci tensor terms in $\RB [h_{\alpha\beta}]$ at order $\mathcal{O}(z^2)$ are absent, which is similar to the contributions of the Weyl tensor term on the brane as shown in eq.~\eqref{eq:zeroWeyl}. 

Lastly, we deal with the extrinsic curvature term associated with $K_{\mt{R}}$ in eq.~\eqref{eq:CVGBbulk}. The unit normal $(t_{\mt{R}})_\alpha$ to the island $\wmB$ embedded on the hypersurface $\mB_{\mt{R}}$ is 
\begin{equation}
(t_{\mt{R}})_\alpha = - \sqrt{h_{zz}(\zB)} \delta^z_\alpha \,. 
\end{equation}
From the definition of the extrinsic curvature, \ie $(K_{\mt R})_{ab} = D_a (t_{\mt{R}})_b $, its trace (in Gaussian normal coordinate) is given by 
\begin{equation}\label{eq:GBK02}
\begin{split}
K_{\mt R} &= -\frac{h^{ab}}{2\sqrt{h_{zz}}} \frac{\partial h_{ab}}{\partial z} \bigg|_{z=\zB}\\
&\approx h^{ab}(\zB) \frac{\tL}{\zB^2}\( 1- \frac{\zB^2}{2\tL^2} \ov{1}{h_{zz}}\) \ov{0}{h_{ab}} + \mathcal{O}(\zB^4) \\
&\approx  \frac{(d-1)}{\tL} \( 1- \frac{\zB^2}{2\tL^2} \ov{1}{h_{zz}}\)- \frac{\zB^2}{\tL^3} \ov{0}{h^{ab}} \ov{1}{h_{ab}}  \\
&\approx \frac{(d-1)}{\tL} -  \frac{\tL}{2} \( \frac{\tK^2}{(d-1)} - \frac{\tR + 2\tR_{ij}\tn^i \tn^j}{(d-2)} \) + \mathcal{O}(\zB^4)\,,
\end{split} 
\end{equation}
where we have recast all geometric quantities as the ones living on the brane in the last line by using eq.~\eqref{eq:rescale} again. Of course, we also find a similar result for $K_{\mt{L}}$. 

Finally, substituting eqs.~\eqref{eq:GBK02}, and \eqref{eq:RB02} into the proposed generalized CV for GB gravity, \ie eq.~\eqref{eq:CVGBbulk},  we can explicitly perform the $z$-integral with lower bound $\zB$ and obtain the leading contributions as
\begin{equation}\label{eq:CVGBbrane02}
\begin{split}
	\mC_\mt{V}^{\rm sub} (\mA_{\mt{L}} \cup \mA_{\mt{R}} )&\approx  \max_{\partial\wmB=\RTbrn}	\left[\frac{2\tL V(\wmB)}{G_{\rm{bulk}}\ell(d-1)} (1+2\lambda f_{\infty} )  \right.\\
&+	\frac{\tL^3 (1-6\lambda f_\infty)}{(d-1)(d-3)G_{\rm{bulk}}\ell}\int_{\wmB} d^{d-1}\sigma \,\sqrt{\tilde{h}}\(  \frac{\tilde{K}^2}{(d-1)}  -\frac{\tilde{R}[\tilde{g}]+ 2\tilde{R}_{ij}[\tilde{g}]\tilde{n}^i\tilde{n}^j}{(d-2)} \)\\
&\left. +\frac{4\lambda f_\infty \tL^3}{(d-1)(d-2)(d-3)G_{\rm{bulk}}\ell}\int_{\wmB} d^{d-1}\sigma \,\sqrt{\tilde{h}}\(  \tK^{ij}\tK_{ij} -\frac{\tK^2}{d-1} + \mathcal{O}(\zB^4) \) \right]\,.
\end{split}
\end{equation}
However, we see this is exactly the expression in eq.~\eqref{eq:CVGBbrane} derived for the induced action on the brane, by noting the relation $\ell' =\frac{d-1}{d-2}\ell$ and $\frac{1}{G_{\rm eff}}= \frac{2\tL}{ d-2} \frac{1+2\lambda f_\infty}{G_{\rm{bulk}}}$. Note that we have counted the double contributions from both sides of the bulk surface $\mB = \mB_{\mt{L}}\cup \mB_{\mt{R}}$ which give rise to the same contributions around the island region.  Therefore, our generalized CV proposal for higher-curvature gravity theory produces consistent results between the bulk gravity theory and brane gravity theory, \ie 
	\begin{equation}\label{eq:equivalence}
\mC^{\rm Island}_{\mt{V}}   \simeq\mC^{\rm sub}_{\mt{V}} \!\( \mA\) \,, 
	\end{equation}
where the maximization of the same functionals over the island region $\wmB$ is considered on both sides as discussed in section \ref{subsec:maximal}.

\subsection{Holographic complexity for $f(\mR)$ gravity}\label{eq:secfR}

In this subsection, we apply the same consistency test with $f(\mR)$ gravity in the bulk to check our proposal. In contrast to the GB theory in the previous subsection, there are extra propagating degrees of freedom in this higher curvature theory \cite{Ezawa:1998ax,Sotiriou:2008rp,Deruelle:2009pu,Deruelle:2009zk}, \ie $f(\mR)$ gravity is properly referred to as a higher derivative theory. We must emphasize the importance of this feature since we saw in the previous section that to properly treat our brane in the limit of zero thickness, the bulk gravity theory should have a good boundary value problem. However, this issue is easily resolved for $f(\mR)$ gravity by recasting it as a scalar-tensor theory -- see below. 

We consider the ($d$+1)-dimensional bulk theory with the action
\begin{equation}\label{eq:actionfR}
	I_{\rm {bulk}}^{f}=\frac{1}{16 \pi G_{\rm{bulk}}} \int d^{d+1} y \sqrt{-g}\left(\frac{d(d-1)}{L^{2}}+f\( \mathcal{R} \) \right) \,.
\end{equation}
In principle, one should consider adding a surface term to this action (\eg see \cite{Dyer:2008hb}) but we will not need to consider the details of this contribution here. Given the above action, it is straightforward to find the equation of motion:
\begin{equation}\label{eq:EoMfR}
f^{\prime}(\mathcal{R}) \mathcal{R}_{\mu \nu}+\left(g_{\mu \nu} \nabla^{\sigma} \nabla_{\sigma}-\nabla_{\mu} \nabla_{\nu}\right) f^{\prime}(\mathcal{R})-\frac{g_{\mu \nu}}{2}\left(f(\mathcal{R})+\frac{d(d-1)}{L^{2}}\right)=8\pi G_{\rm{bulk}}T_{\mu\nu} \,.
\end{equation}
In the absence of matter (\ie with $T_{\mu\nu}=0$), we will assume that this equation is solved by an AdS$_{d+1}$  spacetime whose curvature scale $\tilde L$ is related to $L$ by 
\begin{equation}
	-\frac{d(d-1)}{L^2} =  f( \mR_0 ) + \frac{2d}{\tL^2} f'( \mR_0 )   \qquad{\rm where}\ \  \mR_0  = -\frac{d(d-1)}{\tL^2} \,.
\end{equation}

As emphasized above, $f\( \mR \)$ gravity is a fourth-derivative theory but is classically equivalent to a second-derivative scalar-tensor theory \eg \cite{Sotiriou:2008rp,Deruelle:2009pu}. To be precise, by introducing a scalar field $\Phi$, we can define the classically equivalent scalar-tensor theory with action 
\begin{equation}\label{eq:scalartensor}
\begin{split}
I_{\rm bulk}^{st} = &\frac{1}{16 \pi G_{\rm{bulk}}} \int d^{d+1} y \sqrt{-g}\left(\frac{d(d-1)}{L^{2}}+f\( \Phi \)  + f'(\Phi) \( \mR -\Phi \) \right)\\
& \qquad +\frac{1}{8 \pi G_{\rm{bulk}}} \oint d^{d} x \sqrt{-\tilde{g}}  \, \mK_{\mt{B}} f'(\Phi)\,. 
\end{split}
\end{equation}
Here, we have explicitly introduced the surface term here which produces a well-posed variational principle with Dirichlet boundary conditions \ie $\delta \Phi=0=\delta g_{\mu\nu}$. 
The equation of motion for the scalar field reads  
\begin{equation}\label{eq:onshell}
f''(\Phi) \(  \mR - \Phi \) =0 \,. 
\end{equation}
Imposing the on-shell condition  $\Phi = \mR$ (assuming $f''(\mR) \ne 0$), the action in eq.~\eqref{eq:scalartensor} obviously reduces to eq.~\reef{eq:actionfR} for $f(\mR)$ gravity.  On the other hand, varying the metric yields the field equations 
\beqa
\mR_{\mu\nu} -\frac{1}{2} \mR g_{\mu\nu} &= &\frac{1}{f'(\Phi)} \[ \nabla_{\mu} \nabla_{\nu} f'(\Phi) - g_{\mu\nu} \square f'(\Phi) -\frac{1}{2} g_{\mu\nu} \( \Phi f'(\Phi) - f(\Phi) - \frac{d(d-1)}{L^2} \) \]  
\nonumber\\
&&\qquad\qquad+ \frac{8\pi \Gbulk}{f'(\Phi)}\, T_{\mu\nu} \,.
\eeqa
Upon substituting the on-shell condition $\Phi = \mR$, these equations of motion reduce to the fourth-order equations \eqref{eq:EoMfR} derived by varying the original $f(\mR)$ action. Noting the coefficient associated with the matter stress tensor  $T_{\mu\nu}$, we introduce the ``effective Newton constant" for the $(d+1)$-dimensional scalar-tensor theory as
\begin{equation}
\frac{1}{\hGeff} = \frac{f'(\Phi)}{\Gbulk} \,,
\end{equation}
due to the coupling between gravity and the scalar field $\Phi$. When the matter terms are absent, the bulk spacetime remains the same AdS$_{d+1}$ as above with $\Phi_0=\mR_0$, and in the case,  the ``effective Newton constant'' is actually a constant 
\begin{equation}
\frac{1}{\hGeff} = \frac{f'(\Phi_0)}{G_{\rm{bulk}}} = \frac{f'(\mR_0)}{G_{\rm{bulk}}}\,.
\end{equation}

More generally, we can considering an asymptotically AdS$_{d+1}$ spacetime and one finds that the FG expansion for the Ricci scalar $\mR$ up to the fourth order takes the form 
\begin{equation}
	\mR \[g_{\mu\nu}\]= \mR_0 + \mathcal{O}(\zB^6) \,. 
\end{equation}
by doing a similar calculation to those in the previous subsection. 
Hence with the on-shell condition, we have $\Phi=mR = \mR_0 + \mathcal{O}(\zB^6)$. Further the trace of the extrinsic curvature $\mK_{\mt B}$ at the brane is given by
\begin{equation}
	\mathcal{K}_{\mt B}=\frac{1}{\tL}\left[d+\frac{\tL^{2}}{2(d-1)} \tR+\frac{\tL^{4}}{2(d-1)(d-2)^{2}}\left(\tR_{i j} \tR^{i j}-\frac{d}{4(d-1)} \tR^{2}\right)\right]+\mathcal{O}\left(z_{\mt B}^{6}\right)\,.
\end{equation}
Now integrating out the radial direction in the bulk action in the vicinity of the brane, we obtain the induced gravitational action on the brane as \cite{Pourhasan:2014fba}
\begin{equation}\label{eq:actionfRbrane}
	I_{\rm i n d}=\frac{1}{16 \pi G_{\rm eff}}\int_{\rm brane}\!\!\! d^{d} x \sqrt{-\tilde{g}}\left[\frac{(d-1)(d-2)}{\ell_{\text {eff }}^{2}}+\tR+\kappa_1\left(\tR_{i j} \tR^{i j}-\frac{d}{4(d-1)} \tR^{2}\right)+\mathcal{O}\left(\zB^{6}\right)\right] \,,
\end{equation}
where the various coupling constants  are given by 
\begin{equation}
\begin{split}
\frac{1}{\ell_\mathrm{eff}^2}= \frac{2}{\tilde{L}^2} \(  1 -\right.&\left. \frac{4\pi \tilde{L}\hGeff T_o }{(d-1)} \) = \frac{2}{\tilde{L}^2} \(  1 - \frac{4\pi \tilde{L}\Gbulk T_o }{(d-1)f'(\mR_0)} \)\,,\\
	\frac{1}{G_{\rm eff}}=\frac{2\tL}{(d-2)\hGeff}& = \frac{2\tL}{d-2}\, \frac{f'(\mR_0)}{G_{\rm{bulk}}} \,, \qquad \kappa_1 = \frac{\tL^2}{(d-2)(d-4)} \,.\\
\end{split}
\end{equation}
In producing this result, we have introduced the surface term on either side of the brane, as discussed below eq.~\reef{defE}. 
The induced action on the brane from $f\(\mR\)$ gravity in the bulk is similar to that from Einstein gravity on the bulk in eq.~\eqref{eq:effectiveaction} except for the corrections on the coupling constants from $f'\( \mR_0 \)$.

\subsubsection{Equivalence of the holographic complexities}

Our goal is to show that the relation 
\begin{equation}
\mC_\mt{V}^{\rm sub} \(\mA\) \simeq	\mC^{\rm Island}_\mt{V} \equiv  \max_{\partial\wmB=\RTbrn} \[\frac{\widetilde{W}_{\rm gen}\big(\wmB\big) + \widetilde{W}_{K}\big(\wmB\big)  }{\Geff\,\ell'}   \]\,,
\end{equation}
also holds for $f(\mR)$ gravity in the bulk and its induced gravity on the brane. 
Thanks to the similarity between the induced action in eq.~\eqref{eq:actionfRbrane} and that for Einstein gravity in the bulk, \ie eq.~\eqref{eq:effectiveaction}, it is easy to find that the generalized CV on the brane with this induced gravity theory is given by
\beqa\label{eq:CVfR}
		\mC_\mt{V}^{\rm Island}
		&= &\max_{\partial\wmB=\RTbrn}\left[  \frac{V(\wmB)}{G_{\rm{eff}} \ell'}   \right.\\
		&&+\left. \frac{\tL^2}{2G_{\rm{eff}} \ell'} \int_{\wmB} d^{d-1}\sigma \,\sqrt{\tilde{h}}\(  \frac{\tilde{K}^2}{(d-1)(d-3)}  -\frac{\tilde{R}[\tilde{g}]+ 2\tilde{R}_{ij}[\tilde{g}]\tilde{n}^i\tilde{n}^j}{(d-2)(d-3)}  + \cdots\) \right]\,.
\nonumber
\eeqa

We expect that our proposal can provide the same result as eq.~\eqref{eq:CVfR} by considering the generalized CV in $(d+1)$-dimensional bulk with $f\( \mR \)$ gravity. However, due to the higher-derivative terms, it is much easier to consider the holographic complexity directly in the equivalent scalar-tensor theory \reef{eq:scalartensor} because the gravitational part is only described by the Einstein gravity. Correspondingly, the generalized volume term reduces to a volume term and the K-term simply vanishes. The one subtlety is that we apply our proposal \eqref{eq:defCV} to the scalar-tensor theory with the ``effective Newton constant" $\hGeff= \Gbulk/f'(\Phi)$. However, noticing that $\hGeff$ may be a locally varying quantity on the asymptotically AdS spacetime, we should put the factor $\frac{1}{\hGeff}$ inside the integrals for $W_{\rm gen}, W_{K}$. Then the generalized CV complexity reads 
\begin{equation}\label{eq:CVbulkfR}
\begin{split}
\mC_\mt{V}^{\rm sub} (\mA)
&=\max_{\partial \mB={\mA}\cup {\RT}}	\int_{\mB} d^{d}\sigma\, \sqrt{\det h_{\alpha\beta}} \frac{1}{\hGeff\ell}\left(\alpha_{d+1}\, \frac{\partial \mathbf{L}_{\rm bulk}}{\partial \mR_{\mu\nu\rho\sigma}} n_{\mu} h_{\nu\rho} n_{\sigma}+ \gamma_{d+1}\right) \,,\\
&= \max_{\partial \mB={\mA}\cup {\RT}}  \[ \int_{\mB} d^{d-1}\sigma dz\sqrt{\det h_{\alpha\beta}} \frac{f'(\Phi)}{G_{\rm{bulk}} \ell }  \(  \frac{d}{2}\alpha_{d+1}   +{\gamma}_{d+1}\) \]\,, 
\end{split} 
\end{equation}
where $\mathbf{L}_{\rm bulk} \equiv 16\pi \hGeff\mathcal{L}_{\bulk}$, both $W_K$ and $ W_{\rm bdy}$ vanish due to the absence of higher curvature terms in eq.~\eqref{eq:scalartensor}.  
Substituting the values of $\alpha_{d+1}$ and ${\gamma}_{d+1}$ derived from eq.~\reef{eq:alphagamma}, one can find that the expression in round parentheses reduces to one.  Then extremizing the holographic complexity in the scalar-tensor theory results in 
\beqa
&&\underset{\mB_{\mt{L}},\mB_{\mt{R}}}{\text{ext}} \[ \frac{1}{G_{\rm{bulk}}\ell } \int_{\mB} d^{d-1}\sigma dz\sqrt{\det h_{\alpha\beta}}  \, f'(\Phi) \]
\label{goggle2}\\ 
&& \simeq \frac{2\tL^d  f'(\mR_0) }{G_{\rm{bulk}} \ell} \int_{\wmB} d^{d-1}\sigma \,\sqrt{\det \ov{0}{h}_{ab}}  \left[ \frac{1}{(d-1)\zB^{d-1}} + \frac{1}{(d-3)\zB^{d-3}} \(  \frac{d-2}{2(d-1)^2} K^2  -\frac{R^a_a- \frac{1}{2}R}{2(d-2)} \)  \right]\,, 
\nonumber
\eeqa
where, once again ${\cal B} = \mB_{\mt{L}} \cup \mB_{\mt{R}}$ and $\wmB= \mB_{\mt{L}} \cap \mB_{\mt{R}}$ and we also used the on-shell condition in eq.~\eqref{eq:onshell} and the series expansion $f'(\Phi)=f'(\mR) \approx f'(\mR_0) + \mathcal{O}(\zB^6)$.  
Using the geometric quantities of the brane and noting the maximization over $\wmB$, we can finally obtain the generalized CV for the $f(\mR)$ gravity in the bulk as 
\begin{equation}\label{eq:CVfR02}
\begin{split}
\mC_\mt{V}^{\rm sub} \(\mA \)&= \max_{\partial\wmB=\RTbrn} \left[		 \frac{2\tL f'\( \mR_0\)}{G_{\rm{bulk}} \ell (d-1)}\    \bigg(\ 	V(\wmB) \right.    \\
&\left. \left. 	+ \frac{\tL^2}{2(d-3)}\int_{\wmB} d^{d-1}\sigma \,\sqrt{\tilde{h}}\(  \frac{\tilde{K}^2}{d-1}  -\frac{\tilde{R}[\tilde{g}]+ 2\,\tilde{R}_{ij}[\tilde{g}]\tilde{n}^i\tilde{n}^j}{d-2}  + \mathcal{O}(\zB^4)\) \right)\right] \,,
\end{split}
\end{equation}
Comparing eqs.~\eqref{eq:CVfR} and \eqref{eq:CVfR02}, we also find the equivalence between the holographic complexity derived from $f(\mR)$ gravity and its induced gravity theory on the brane, \ie 
\begin{equation}
\mC_\mt{V}^{\rm Island}\simeq \mC_\mt{V}^{\rm sub} \(\mA\)\,,
\end{equation}
where we have used the relations $\ell'=\frac{d-1}{d-2}\ell$ and $\frac{1}{G_{\rm eff}}= \frac{2\tL}{d-2} \frac{f'(\mR_0)}{G_{\rm{bulk}}}$.
Once again, this equivalence supports that our proposed holographic complexity for higher-derivative gravity theory produces consistent results.

%% file: sections/Discussion.tex

As discussed in section \ref{subsec:maximal}, there is an interesting identification between the island rule \eqref{eq:islandformula} for the brane perspective and the RT prescription \eqref{eq:island2} for the bulk perspective in the doubly holographic model of \cite{Chen:2020uac,Chen:2020hmv}. One feature is that the RT surfaces in eq.~\reef{eq:island2} are extremized in two stages: first, one finds surfaces that are extremal everywhere aware from the brane, and second, the intersection of the RT surfaces with the brane is extremized. The latter corresponds to finding the quantum extremal surface in the island rule \eqref{eq:islandformula}. The on-shell bulk surfaces found in the first step describe the leading contributions to the entanglement entropy in the large $N$ limit of the boundary CFT, for different candidate quantum extremal surfaces. These contributions may be divided into two classes: various geometric contributions corresponding to terms of the Wald-Dong entropy \cite{Wald:1993nt,Iyer:1994ys,Jacobson:1993vj,Dong:2013qoa} coming from the various gravitational interactions induced in the brane theory by the CFT,\footnote{Of course, these must be combined with the brane contribution in eq.~\reef{eq:island2} to produce the full Wald-Dong entropy  of the effective gravity theory on the brane \cite{Chen:2020uac}.} and the quantum contributions appearing as $S_\mt{QFT}$ in the island rule \eqref{eq:islandformula}. Of course, the first set of contributions comes from integrating the bulk area of the RT surface near the brane, while the second set comes from the bulk region far from the brane.\footnote{These include both UV contributions from near the asymptotic boundary and IR contributions from deep in the AdS bulk.}  

As discussed in section \ref{subsec:maximal}, there seems to be a direct parallel between the above analysis of the holographic entanglement entropy and of the holographic complexity using the CV proposal. Hence beginning with the subregion complexity=volume proposal \reef{eq:CV-notboth} in the bulk,\footnote{Or alternatively, eq.~\reef{eq:CV-both} for a DGP brane.} we arrive at the following description of the complexity from the brane perspective:
\begin{equation}\label{eq:CVequalsisland2}
\mC_{\mt{V}}^{\rm sub} ( \mA)  = \max_{\partial\wmB=\RTbrn} \[ \frac{\widetilde{W}_{\rm gen}( \wmB) + \widetilde{W}_K (\wmB) }{G_{\rm eff}\ell'}  +\mC_{\mt{QFT}}(\mA \cup \wmB) \]  \,,
\end{equation}
where the geometric contribution is given by eqs.~\eqref{eq:fixW}  and \eqref{eq:fixWK}. Focusing on this geometric contribution, this result leads us to propose eqs.~\reef{eq:ourporposal} and \reef{eq:details} as the extension of the CV proposal for holographic complexity in higher curvature theories. Our experience with the Wald-Dong entropy suggests that $W_K$ provides an infinite series of corrections involving higher powers of the extrinsic curvature \cite{Dong:2013qoa}, and eq.~\reef{eq:details} only presents the first $K^2$ term in this series. Further, in section \ref{OTbrain}, we noted that the $K$-term in eq.~\eqref{eq:defineWK} was chosen for its simplicity and the similarity to the form of the $K$ corrections in the Wald-Dong entropy, but we cannot rule out the possibility that it involves more complicated contractions than that in eq.~\eqref{eq:defineWK}. 

\begin{figure}[htbp]
	\centering\includegraphics[width=0.5\textwidth]{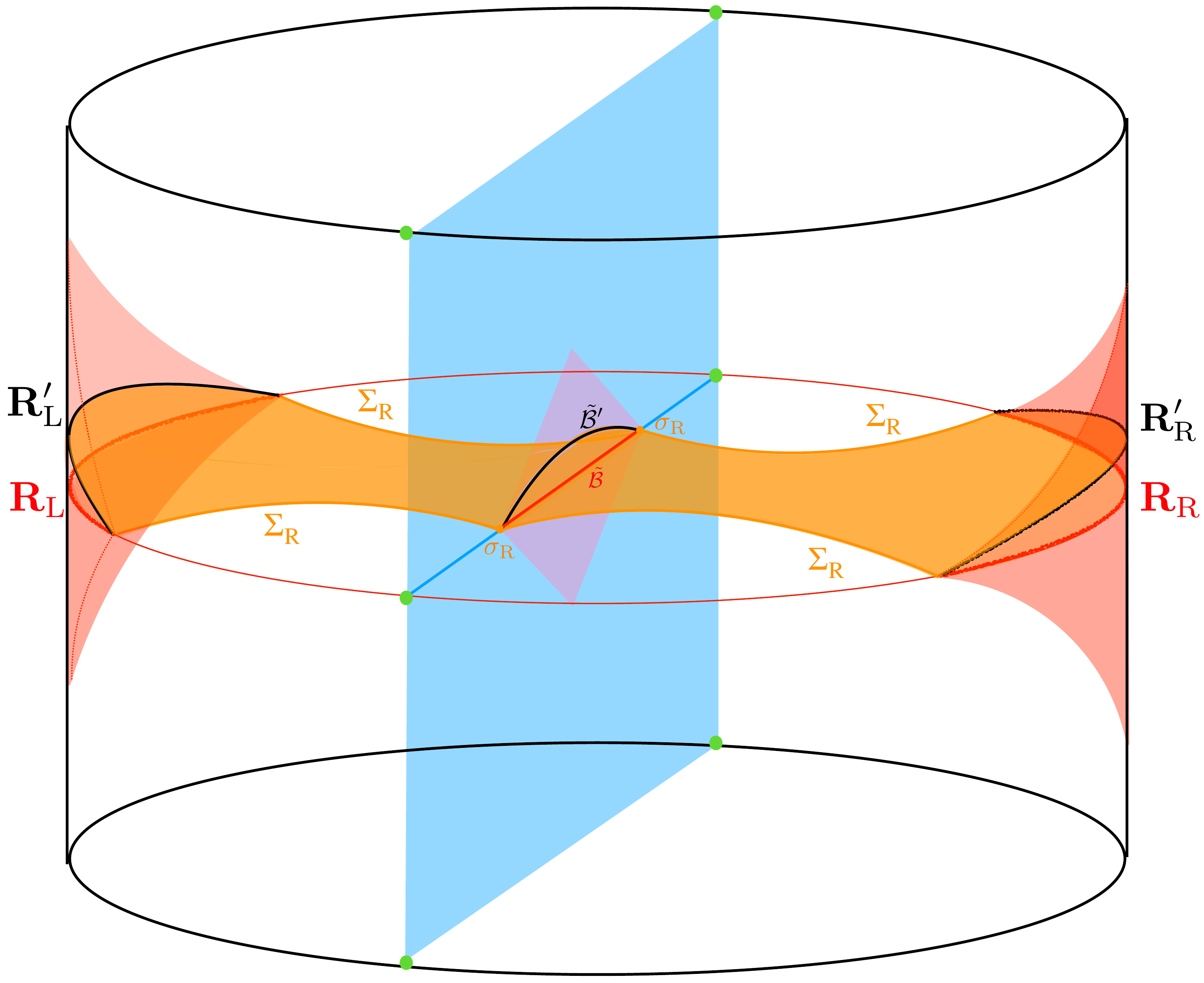}
	\caption{Different boundary subregions, $\mA=\mA_{\mt{L}} \cup \mA_{\mt{R}}$ and $\mA'=\mA'_{\mt{L}} \cup \mA'_{\mt{R}}$ with the same boundaries, \ie $\partial\mA =\partial\mA'$. The entanglement entropy and the RT surface remains the same for both subregions. However, the extremal surfaces $\mB$ and $\mB'$ (denoted by the orange regions) are different and hence they produce different islands $\wmB$ and $\wmB'$ on the brane (represented by the blue slice). The QES on the brane is unchanged and hence $\partial\wmB=\partial\wmB' =\RTbrn$. The red shaded regions on the asymptotic boundary represent the causal domain of $\mA$  ($\wmB$). The subregion $\mA'$ may be any spacelike surface in this causal domain. Similarly, $\wmB'$ will always lie within the causal domain of the brane (denoted by the pink region). 
		\label{bdy-surface}}
\end{figure}

Perhaps, equally interesting in eq.~\reef{eq:CVequalsisland2} are the `quantum' contributions coming from integrating the bulk volume of the extremal surface far from the brane. These contributions can play an important role in determining the geometry of $\wmB$. Recall that the boundary of the extremal surface consists of $\partial\mB=\mA\cup\RT$. Hence the profile of $\mB$ depends on the full details of the geometry of the boundary subregion $\mA$. Hence any two $\mA$ and $\mA'$ with $\partial\mA =\partial\mA'$ yield the same RT surface $\RT$ and the same quantum extremal surface $\RTbrn$ on the brane, but these different choices will produce different island surfaces $\wmB$ -- see figure \ref{bdy-surface}. Of course, this reflects the fact that the holographic complexity is sensitive to the details of the state that are not captured by the corresponding entanglement entropy. 

A simple observation is that  the holographic CV calculation picks out a special time slice on the brane (\ie $\wmB$) for the island, in contrast to the corresponding entanglement entropy calculation  which only fixes the  boundary of the islands (\ie the quantum extremal surface). It would be interesting to explore how $\wmB$ is deformed by making variations of the subregion $\mA$ on the asymptotic AdS boundary, or perhaps by the insertion of extra operators in this subregion. While in principle these deformations could fill the causal domain of some canonical time slice with boundary $\RTbrn$, our intuition is that generally, such variations will only produce perturbatively small deformations of $\wmB$. If one examines the FG expansion \reef{eqx1FG} for embedding surface near the brane more closely, one finds
		\begin{equation}
		x^i \( z, \sigma^a \) = \ov{0}{x^i}\(\sigma^a \)  +\ov{1}{x^i}\(\sigma^a \) +\cdots+ \frac{z^d}{L^d}  \(   \ov{d}{x^i} \( \sigma^a\)  + \ov{d}{y^i} \( \sigma^a\)  \log\( \frac{z}{L}  \) \)+ \mathcal{O}\(\frac{z^{d+1}}{L^{d+1}}\)\,. 
		\end{equation}
The coefficients $\ov{n}{x}{}^i$ with $n<d$ are completely determined by the boundary profile $\ov{0}{x}{}^i$ and the boundary metric $\ov{0}{g}{}_{\! ij}$, \eg see \cite{Hung:2011ta}. The second independent coefficient in this expansion is $\ov{d}{x}{}^i$. This is precisely the coefficient that is determined by the infrared physics and the shape of $\mA$ and so naively, its contributions on the brane are suppressed by the power $(z_\mt{B}/L)^d\ll1$ in the regime of interest. 

The above expansion  also resolves a puzzle with eqs.~\reef{eq:CV-notboth} and \reef{eq:CVequalsisland2}. In the latter equation, the brane perspective seems to treat $\mC_{\mt{QFT}}$ as a higher-order term of the expansion in $G_{\rm eff}$. However, both contributions arise at the same order in the $G_\mt{N}$ expansion in the bulk. There is no contradiction because the quantum corrections from the boundary CFT are enhanced by a power of the central charge $c_\mt{T}\sim L^{d-1}/G_\mt{N} \sim L^{d-2}/G_{\rm eff}$ -- where we use eq.~\reef{eq:Gd} in the latter. However, the effect of the quantum contribution $\mC_{\mt{QFT}}$ can still be suppressed in the expansion on the brane in terms of powers of $z_\mt{B}/L\sim L/\ell_\mt{B}$.

A fascinating aspect of the second term in eq.~\reef{eq:CVequalsisland2} is that while this contribution has a geometric origin in our bulk calculations, it is interpreted as a quantum contribution from the brane perspective, \ie it is associated with the quantum fields on $\mA\cup\wmB$. The interpretation follows the parallel with the holographic entanglement and the appearance of $S_\mt{QFT}( \bdyReg \cup \islands )$ in the island rule \reef{eq:islandformula}. Of course, it points to an improved version of our generalized complexity=volume proposal \reef{eq:ourporposal} of the form
\begin{equation}\label{corrected}
\mC_\mt{V}(\mA)=\max_{\partial\mB=\mA} \[  \frac{W_{\rm gen}(\mathcal{B}) + W_K(\mathcal{B})}{G_\mt{N}\,\ell} + \mC_{\mt{bulk}}  \] \,,
\end{equation}
where $\mC_{\mt{bulk}}$ represents the contribution from the quantum field state in the bulk. This would be analogous to the appearance of quantum corrections in the holographic entanglement entropy \cite{Faulkner:2013ana,Engelhardt:2014gca}. Of course, such additional contributions have long been expected because the CV proposal \reef{eq:CVpure} has the form of a saddle point approximation of some more complete calculation. While eq.~\reef{eq:CVequalsisland2} is the first instance where these quantum corrections can be explicitly calculated, unfortunately, our doubly holographic model does not indicate what quantum calculation yields these contributions. Of course, it would be interesting to further investigate the properties of $\mC_{\mt{QFT}}$ in eq.~\reef{eq:CVequalsisland2} to gain further insight into this question. 

In this vein, one immediate observation from examining eq.~\reef{eq:CVequalsisland2} is the tension between the maximization and the naive association of $\mC_\mt{QFT}$ with circuit complexity -- or rather circuit depth. That is, if we associate $\mC_\mt{QFT}$ with the size of the quantum circuit needed to prepare the QFT state on the corresponding region (along the lines studied in, \eg \cite{Jefferson:2017sdb,Chapman:2017rqy}) then the complexity follows from minimizing this quantity rather than maximizing. One simple resolution would be to consider our analysis with a Euclidean (rather than a Minkowski) signature. Then the CV conjecture \reef{eq:CVpure} would correspond to minimizing the volume of the bulk surfaces and this minimization would naturally be inherited by the generalized proposal in eq.~\reef{eq:CVequalsisland2} or \reef{corrected}. This tension may suggest that $\mC_\mt{QFT}$ should instead be associated with an alternative interpretation of holographic complexity, \eg optimization of path integrals \cite{Caputa:2017urj,Caputa:2017yrh},
``quantum circuits'' based on path-integrals \cite{Takayanagi:2018pml} or using the equivalence of bulk and boundary symplectic forms \cite{Belin:2018fxe,Belin:2018bpg,Belin:2020oib}. Our doubly holographic model may also provide an interesting new forum to study these approaches.

\subsection*{General higher curvature theories}

While we are proposing that the generalized expressions for holographic complexity in eqs.~\reef{eq:ourporposal} and \reef{eq:details} should apply for general theories of higher curvature gravity, we only applied our consistency tests in section \ref{sec:GB} to two very specific theories. The feature that distinguished these theories was their boundary value problem. Namely, Gauss-Bonnet gravity can be solved with standard boundary conditions applied to the metric, while $f(\mR)$ gravity could be expressed in a form (\ie as a scalar-tensor theory) where the boundary conditions had a simple form. We note however that this limitation was because of issues in dealing with infinitely thin brane in higher curvature theories. Hence while this is a limitation of the doubly holographic model, we do not believe that it limits the applicability of our generalized proposal for holographic complexity.
Certainly, the induced gravitational theories on the brane are outside this limited class of higher curvature theories.

However, we must admit that there are aspects of our consistency tests in section \ref{sec:GB} that deserve further consideration. For example, one should better understand the appearance of the ``effective Newton constant" in the generalized volume for $f(\mR)$ gravity. For the Gauss-Bonnet theory, it would be interesting to understand how to derive the expression for $W_{\rm bdy}$ in eq.~\reef{eq:GB_Wbdy}  from the surface terms added to the gravitational action on either side of the brane.
 
We hope our generalized extension of the CV proposal will encourage further investigations of holographic complexity in higher curvature gravity models. Many studies of the CV proposal for higher derivative gravity (\eg \cite{Alishahiha:2017hwg,An:2018dbz,Nally:2019rnw, Jiang:2019kks}) only consider the volume term. Therefore it will be interesting to explore the differences between the CV and our generalized CV approaches in various settings.

To close here, let us add that there is another interesting discrepancy in our approach which deserves further study. Setting aside the doubly holographic model and considering standard AdS holography for a moment, we observe that one finds logarithmic divergences in evaluating the boundary counterterms and the holographic entanglement entropy when the boundary dimension $d$ is even. Of course, these divergences are related to the conformal anomaly of the boundary CFT. However, in evaluating the extremal volume for the holographic complexity, one finds that there are logarithmic divergences when $d$ is odd. As a result, in the analysis of the doubly holographic model, one finds that one can account for the log divergences in the entanglement entropy (coming from the bulk region near the brane) by straightforwardly applying the Wald-Dong entropy to the induced gravitational action on the brane \cite{Chen:2020uac}. In contrast, there is no such match between the logarithmic divergences in the CV complexity in the bulk and the geometric contributions in our generalized complexity \reef{corrected} on the brane (for odd $d$). Similarly, applying our geometric formula to the logarithmically divergent terms in the induced action naively yields contributions which do not appear in the CV complexity (for even $d$). In either case, one can adopt an approach where these logarithmic terms are treated separately. However, an alternative may be that the boundary between the geometric gravitational contributions and the quantum contributions is different for the generalized CV complexity in eq.~\reef{corrected}, than say, for holographic entanglement entropy. This is certainly an issue that deserves further consideration.

While the above issue arises for all values of $d$ when calculating corrections to sufficiently high orders, let us add that it is immediately apparent in our analysis in section \ref{OTbrain} for lower dimensions, \eg the coefficients in eq.~\reef{eq:alphagamma} diverge for $d=2$ or 3. It arises there because the logarithmic divergence appears in the leading or first subleading contribution. We provide a detailed examination of these two cases in appendix \ref{app:lower}. However, we emphasize again that the same issue arises in higher dimensions but only in higher-order contributions.

\subsection*{Mutual complexity and island complexity}

Much of our analysis focused on identifying the geometric terms in eq.~\eqref{eq:CVequalsisland2} by looking at the contributions arising from the region near the brane, \ie the leading terms in the limit ${\zB}/{\tilde{L}} \to 0$. However, we should recall that the quantum term $\mC_{\mt{QFT}}(\mA \cup \wmB)$ also includes the UV divergent terms associated with the cut-off surface near the asymptotic AdS boundary. These are less interesting for our purposes and so we point out that they can be eliminated by considering the mutual complexity, \eg \cite{Couch:2018phr,Agon:2018zso,Caceres:2019pgf,Ruan:2020vze}
\begin{equation}
\Delta \mC_{\mt{V}}^{\rm sub} =\mC_{\mt{V}}^{\rm sub} \( \mA_{\mt{L}} \) + \mC_{\mt{V}}^{\rm sub} \( \mA_{\mt{R}} \) -\mC_{\mt{V}}^{\rm sub} \( \mA_{\mt{L}} \cup \mA_{\mt{R}} \)\,.
\end{equation}
The UV divergent terms, which only depend on the boundary geometry of $\mA_\mt{L}$ and $\mA_\mt{R}$, cancel in this combination of complexities, leaving a UV finite quantity.

We also remark that the transition between the no-island phase to the island phase can also be diagnosed by the above mutual complexity.
In particular, the latter vanishes in the no island phase, in which the bulk RT surfaces are disconnected phase -- see figure \ref{fig:simple}. For the island phase, the mutual complexity jumps to a large negative value. In fact, we expect that this is dominated by the island contribution, \ie 
\begin{equation}
\Delta \mC_{\mt{V}}^{\rm sub}  \simeq -\, \mC_{\mt{V}}^{\rm Island} + \cdots = -\,  \frac{\widetilde{W}_{\rm gen}( \wmB) + \widetilde{W}_K (\wmB) }{G_{\rm eff}\ell'}\bigg|_{\wmB_\mt{ext}}  +\cdots\,.
\end{equation}
Even though the entanglement entropy is continuous at the transition between these two phases, the complexity of the island state is much larger than that of the no island state. This reflects the fact that one is able to reconstruct the island on the brane from the asymptotic boundary state. Of course, similar discontinuities in the mutual complexity are seen in more conventional holographic settings, \eg \cite{Ben-Ami:2016qex,Couch:2018phr,Caceres:2018blh,Caceres:2019pgf,Auzzi:2019vyh,Braccia:2019xxi,Sato:2019kik,Bhattacharya:2020uun}, but it would interesting to further understand the implications for quantum extremal islands.

\subsection*{Length scale in holographic complexity}

Both the holographic CV proposal \eqref{eq:CVpure} and our proposed generalization \eqref{eq:ourporposal} involve an undetermined length scale $\ell$. In most previous studies, \eg \cite{Carmi:2016wjl,Couch:2016exn,Ben-Ami:2016qex,Carmi:2017jqz,Swingle:2017zcd,Chapman:2018dem,Chapman:2018lsv,Fu:2018kcp,Flory:2018akz,Bernamonti:2020bcf,Chen:2020nlj,Sarkar:2020yjs,Alishahiha:2015rta,Abt:2017pmf,Bakhshaei:2017qud,Abt:2018ywl,Caceres:2018luq,Chapman:2018bqj,Agon:2018zso,Chen:2018mcc,Caceres:2018blh,Bhattacharya:2018oeq,Cooper:2018cmb,Caceres:2019pgf,Bhattacharya:2019zkb,Karar:2019bwy,Sato:2019kik,Auzzi:2019fnp,Braccia:2019xxi,Lezgi:2019fqu,Auzzi:2019mah,Ling:2019ien,Chakrabortty:2020ptb,Auzzi:2019vyh,Balushi:2020wjt,Cai:2020wpc},\footnote{See \cite{Couch:2018phr} for a more elaborate definition of this scale in the case of holographic black holes.} this length scale is simply chosen to be the AdS curvature scale. However, our analysis was simplified by leaving $\ell$ undetermined, and in particular, we found a simple relation \reef{newel} between the scales associated with the holographic complexity in the bulk and on the brane.  The AdS radius of the induced gravity on the brane, \ie $\ell_\mt{B} \approx {{L}^2}/{\zB}$, is more or less independent of the bulk radius ${L}$, \ie the relation depends on the brane tension as shown in eq.~\reef{curve2}. If one demands to identify the length scale in the complexity proposals to the AdS radius for the different gravity theories, the generalized CV for boundary subregion and the island are given by 
 \begin{equation}
 \begin{split}
  C_{\mt{V}}^{\rm sub}  &\equiv \max_{\mB} \[  \frac{1}{\Gbulk\,{L}} \( W_{\rm gen}(\mB) + W_K(\mB)  \)  \] \,,\\
  C^{\rm Island}_\mt{V}&\equiv \max_{\wmB} \[  \frac{1}{\Geff\,\ell_\mt{B}} \( \tilde{W}_{\rm gen}(\wmB) + \tilde{W}_K(\wmB)  \)  \] \,,
 \end{split}
 \end{equation}
and the two expressions do not agree, \ie $C_{\mt{V}}^{\rm sub}/C^{\rm Island}_\mt{V} \simeq \ell_\mt{B}/L$. Rather one would have to introduce an additional `penalty factor' to produce the desired  equivalence, \ie
 \begin{equation}
  C_{\mt{V}}^{\rm sub}  \simeq\mathcal{P} \, C^{\rm Island}_\mt{V} + \cdots \qquad{\rm with}\ \  \mathcal{P}=\frac{d-2}{d-1}\,\frac{\ell_\mt{B}}{L}\,.
 \end{equation}
In contrast to the simple relation in eq.~\reef{newel}, this additional factor has a complicated dependence on the physical parameters of the underlying theory.

\subsection*{Maximal condition for holographic complexity}

As we have stressed, the CV conjecture \reef{eq:CVpure} and our generalized proposal \reef{eq:ourporposal} relies on maximizing the corresponding geometric functional on bulk hypersurfaces $\cal B$ with the appropriate boundary condition $\partial {\cal B} = \mA \cup \Sigma_\mA$. However, we only explicitly use the local equations, \ie $\frac{\delta \mC_{\mt{V}}}{\delta X^\mu}= 0$ to find the extremum. For eq.~\reef{eq:CVpure}, we are guaranteed that the extremal volume will be a maximum. However, with our generalization \reef{eq:ourporposal}, we are no longer guaranteed that the corresponding geometric functional will reach a maximum in situations where the higher curvature contributions become important. That is, the solutions of the extremizing equation may be a maximum, a minimum, or a saddle point. Maximizing  the holographic complexity further requires a necessary condition for the generalized CV functional to be a local maximum, \ie 
 \begin{equation}\label{eq:CVstability}
 \delta^2 \mC_{\mt{V}} \le 0\,,
 \end{equation}
 where the variation is defined with respect to perturbations of the extremal surface $\mB$.  Although, this condition is not necessary for the derivation of the results in this paper,  it is still interesting to explore the meaning of this constraint on second variations of generalized complexity. From the viewpoint of holographic entanglement entropy $ S_{\mt{EE}}$, its second variations (with respect to  deformations of the entangling surface) are also constrained by strong stability, \ie $ \delta^2 S_{\mt{EE}}\ge 0$, due to the fact the RT surface is a local minimum of its area. Similar strong stability should also be imposed on the generalized entropy $S_{\rm gen}$ -- see \cite{Wall:2012uf,Engelhardt:2019hmr} for more discussion. It is remarked that strong stability is a nontrivial constraint independent of its extremality condition. As an important application, the second variation plays a crucial role in defining quantum null energy conditions \cite{Bousso:2015mna,Bousso:2015wca}. So we expect that there will be interesting applications of the stability condition \eqref{eq:CVstability} for holographic complexity.

\subsection*{Generalized first law for causal diamonds}

By applying Wald's Noether charge formalism \cite{Wald:1993nt,Iyer:1994ys}, the authors in \cite{Jacobson:2015hqa,Jacobson:2018ahi} derived an extended first law of causal diamond mechanics in Einstein gravity 
\begin{equation}
\delta H_{\zeta}^{\mathrm{matter}}=-\frac{\kappa}{8 \pi G_{\mathrm{N}}}\left[\delta A-k\delta V\right]\,,
\end{equation}
where $H_{\zeta}^{\mathrm{matter}}$ is the matter Hamiltonian associated with the flow generated by the conformal Killing vector $\zeta$ on the causal diamond, $A$ is the area of the edge $\partial \Sigma $, and $k$ denotes the extrinsic curvature of $\partial \Sigma$ embedded in the maximal slice. Connections to the first law of holographic complexity were also developed in \cite{Belin:2018bpg,Bernamonti:2020bcf,Sarkar:2020yjs}. Furthermore, it was extended to higher-curvature gravity in \cite{Bueno:2016gnv} as 
\begin{equation}
 \delta H_{\zeta}^{\mathrm{matter}}=-\frac{\kappa}{2 \pi G_{\mathrm{N}}}\, \delta S_{\rm Wald} \Big|_{W} + \int_{\partial \Sigma} \delta C_{\zeta} \,,
\end{equation}
where $\delta C_{\zeta} =0$ are the linearized equations of higher derivative theory and the Wald entropy evaluated on bifurcation surface $\partial \Sigma$ varies while keeping the generalized volume $W$ fixed.\footnote{Here the generalized volume $W$ differs from ours in eq.~\eqref{eq:details} because the former has a constant term depending on the couplings of the higher derivative theory and it is normalized to be the regular volume for any higher-curvature gravity when evaluated in AdS background. The simplicity from our complexity formula is that the relevant coefficients only depend on the dimension of theory.} Considering that our proposal suggests a new term $W_{K}$ depending on the extrinsic curvature, it would be interesting to generalize the first law of causal diamond mechanics by connecting the Wald-Dong entropy and our generalized volume.

\subsection*{Generalizing complexity=action?}

In the context of holographic complexity, the complexity=action (CA) conjecture \cite{Brown:2015bva,Brown:2015lvg} and its subregion version \cite{Carmi:2016wjl} have also been widely studied. Generalizing our work to consider the CA proposal in the framework of our doubly holographic model is an obvious future direction. However, in contrast to the CV proposal, the CA approach already includes the corrections from higher-curvature terms due to the explicit dependence of action on these terms. So the real question to verify is whether the subregion-CA proposal in bulk theory produces the same complexity for the induced gravitational theory on the brane, \ie does one find
\begin{equation}
  \mC_{\mt{A}}^{\rm sub} \simeq \mC_{\mt{A}}^{\rm Island} +\cdots \quad  ? 
\end{equation}
If this is not the case, it may imply the need to consider a modified CA approach for higher-curvature gravity theory. Of course, subtlety is that surface and joint terms play a very important role in the CA approach \cite{Lehner:2016vdi}, and determining the corresponding terms for higher curvature theories is quite
demanding, \eg \cite{Cano:2018aqi,Cano:2018ckq,Chakraborty:2018dvi,Jiang:2018sqj}. Let us also note that the $\mC_{\mt{A}}^{\rm sub}$ approach has already been studied in the literature, \eg \cite{Chapman:2018bqj,Braccia:2019xxi,Sato:2019kik}, but the extension to the present context is not obvious from these results.

%% file: sections/lowerD23.tex

In the discussion section, we commented on a discrepancy in our analysis related to logarithmic divergences in the CV complexity and the induced gravity action on the brane. In particular, for odd $d$, the CV complexity in the bulk contains a logarithmic contribution but the latter is not generated by our generalized complexity proposal \reef{eq:new} applied to the corresponding brane action. Similarly for even $d$, applying our geometric formula to the logarithmically divergent terms in the induced action naively yields contributions that do not appear in the CV complexity. Further, this issue becomes immediately evident in lower dimensions, where the logarithmic divergences appear as the leading or first subleading contributions. Explicitly, one can see that our proposal to the generalized CV for a $d$-dimensional gravity theory
\begin{equation}\label{eq:pro}
\mC_\mt{V} \( \mA \)=\max_{\partial \mB={\mA}\cup {\RT}} \[  \frac{W_{\rm gen}(\mathcal{B}) + W_K(\mathcal{B})   }{G_{\mt{N}}\ell} \] \,, 
\end{equation}
is only valid for $d>3$ due to superficial divergences in the coefficients 
\begin{equation}\label{goggle}
\alpha_d=\frac{2 (d-4)}{(d-2)(d-3)}  \,,\qquad \gamma_d=\frac{2 }{(d-2)(d-3)} \,,\qquad A_{d}= \frac{4(d-4)}{(d-2)^2(d-3)}\,,
\end{equation}
when $d=2$ or $d=3$. (Recall that $\beta_d=0$ for all dimensions.) 	Hence in this appendix, we examine this issue by revisiting our analysis in section \ref{OTbrain} for lower-dimensional gravity theories. 

\subsection{Three-dimensional brane} \label{subsec:d3}

We begin here with the case of $d=3$.\footnote{This is the case of three-dimensional gravity, \ie $d=2$ in eqs.~\reef{eq:ourporposal} and \reef{eq:details}.} It is obvious that there is a problem for the subleading contributions in eq.~\eqref{eq:full_subCV} coming from integrating the volume of the extremal surface in the vicinity of the three-dimensional brane. The divergence in the corresponding coefficients is a signal of the appearance of logarithmic terms. Explicitly, performing the $z$-integral for $d=3$, we find that the subregion complexity for four-dimensional bulk gravity reads
\begin{equation}\label{eq:full_subCVd3}
\begin{split}
&\mC^{\rm{sub}}_\mt{V} ( \mathbf{R} )\equiv  \max_{\partial \mB={\mA}\cup {\RT}} \[ \frac{V\( \mathcal{B}_{\mt{L}}\)+V\( \mathcal{B}_{\mt{R}}\)  }{\Gbulk \ell}  \] \\
&\simeq  \frac{2L^2}{\Gbulk \ell} \int_{\wmB} d^2\sigma \int_{\zB}dz\,\sqrt{\det \ov{0}{h}} \( \frac{L}{z} \)^3\( 1-\frac{z^2}{8} K^2 + \frac{z^2}{2L^2} \ov{0}{h^{ab}}\ov{1}{h_{ab}} + \cdots \) \\
&\simeq \frac{L V(\wmB)}{\Gbulk \ell} +  \log\( \frac{\ell_{\mt{IR}}}{\zB}\)	\frac{L^3}{\Gbulk \ell}\int_{\wmB} d^{2}\sigma \,\sqrt{\det \tilde{h}}\(  \frac{\tilde{K}^2}{4} -\frac{1}{2}\tilde{R}- \tilde{R}_{ij}\tilde{n}^i\tilde{n}^j\) +\mathcal{O}(\zB^0)  \,,
\end{split}
\end{equation}
where $\ell_{\mt{IR}}$ is some scale from deep in the bulk which makes the argument of the logarithmic term dimensionless. Hence the leading term in $\mC^{\rm{sub}}_\mt{V} \( \mathbf{R} \)$ still yields the expected volume contribution for the brane gravity, \ie $V(\wmB)/(\Geff \ell')$ with $\ell'= 2\ell$ and $\Geff= \Gbulk/(2L)$ as before.
However, the proposed functional for the generalized CV proposal must be modified at higher orders to match the logarithmic divergence 
\begin{equation}\label{hot}
\mC_{\mt{V},d=3}^{\log}( \wmB ) =\log\!\( \frac{\ell^2_{\mt{B}}}{L^2}\)\,	\frac{L^2}{4\Geff\, \ell'}\int_{\wmB} d^{2}\sigma \,\sqrt{\det \tilde{h}}\(  \frac{\tilde{K}^2}{2} -\tilde{R}- 2\tilde{R}_{ij}\tilde{n}^i\tilde{n}^j\) \,,
\end{equation} 
where we have substituted $\ell_\mt{B}=L^2/z_\mt{B}$ and made the simple choice $\ell_{\mt{IR}}=L$. Recall that $\ell_\mt{B}$ and $L$ correspond to the AdS curvature and the UV cutoff scales, respectively, in the effective theory on the brane \cite{Chen:2020uac,Chen:2020hmv}. Then, we arrive at the generalized CV expression for the induced gravity on the three-dimensional brane,
\begin{equation}\label{eq:d302}
\mC_{\mt{V}}^{\rm sub} (  \mathbf{R}  )\simeq \mC^{\rm{Island}}_\mt{V,d=3}  \equiv  \max_{\partial\wmB=\RTbrn} \[ \frac{V(\wmB) }{G_d\ell'} +\mC_{\mt{V},d=3}^{\log} ( \wmB ) \] 
\end{equation}
where the logarithmic term is explicitly shown in eq.~\eqref{hot} and denotes the contributions from curvature-squared terms in the gravitational action
\reef{eq:effectiveaction}.

Following the approach in the main text, it is straightforward to extend eqs.~\reef{eq:defineWgen} and \reef{eq:defineWK} to the present case if we allow for logarithmic coefficients. Explicitly, we obtain 
\begin{equation}\label{eq:d3}
\begin{split}
\mC^{\rm{Island}}_\mt{V,d=3}&=\frac{1}{G_{\rm eff}\ell' } \int_{\wmB} d^{2}\sigma \sqrt{ \tilde{h}} \left[ \left(1+\log\( \frac{\ell^2_{\mt{B}}}{L^2} \) -\log\( \frac{\ell^2_{\mt{B}}}{L^2} \)\, \frac{\partial \mathbf{L}_{\rm eff}}{\partial \tR_{ijkl}} \tn_{i} \tilde{h}_{ik} \tn_{l}\right) \,, \right.\\
  &\left.-  2\log\( \frac{\ell^2_{\mt{B}}}{L^2}\)  \frac{\partial^2 \mathbf{L}_{\rm eff}}{\partial \tR_{ijkl}\partial \tR^{mnop}}  \tK_{jl}\tilde{h}_{ik} \tK^{np} \tilde{h}^{mo} \right] \,.
\end{split}
\end{equation}
That is, we are using the same functional $\widetilde{W}_{\rm gen} + \widetilde{W}_{K}$ as before but with new coefficients 
\begin{equation}
\alpha_3 =-\log\( \frac{\ell^2_{\mt{B}}}{L^2} \)  \,, \quad  \gamma_3=1+\log\( \frac{\ell^2_{\mt{B}}}{L^2} \) \,,\qquad A_{3}=-2\log\( \frac{\ell^2_{\mt{B}}}{L^2}\) \,,
\end{equation}
for a general curvature-squared gravity theories in three dimensions.

We emphasize that we included the first subleading contributions in eq.~\eqref{eq:full_subCV} and so the issue of the logarithmic divergence in the holographic complexity became manifest for $d=3$. However, the same issue will arise for any odd $d$, \ie with an even dimension in the bulk. Carrying out the same calculations to a sufficiently high order will reveal an extra logarithm in the holographic complexity. In particular, with $d=2n+1$, one should only apply eqs.~\reef{eq:ourporposal} and \reef{eq:details} for the generalized CV proposal for higher curvature interactions up to $R^{2n-1}$. It will be possible to include the $R^{2n}$ interactions if one adds an extra contribution with a logarithmic coefficient, as in eq.~\reef{eq:d302}.
It would be interesting to examine this issue in greater detail in higher dimensions.

\subsection{Two-dimensional brane}

Now turning to the case of $d=2$,\footnote{This is the case of two-dimensional gravity, \ie $d=1$ in eqs.~\reef{eq:ourporposal} and \reef{eq:details}.} we expect to find a logarithmic divergence in the induced action which is not reflected in the holographic complexity. Furthermore, we should stress that the generalized CV for $d=2$ is more subtle because the usual relations $\ell'= \frac{d-1}{d-2}\ell$ and $\Geff=(d-2)\Gbulk/(2L)$ break down for this dimension. 

First of all, we recall the FG expansion for the metric with a three-dimensional bulk becomes
\begin{equation}
g_{ij}( z, x^i) =  \ov{0}{g}_{ij} \( x^i \) +   \frac{z^2}{\delta^2} \(  \ov{1}g_{ij} (x^i)  + f_{ij}(x^i) \log \( \frac{z}{L} \)    \)  + \cdots \,. 
\end{equation}
where the subleading term $ \ov{1}g_{ij} (x^i)  $ is not completely fixed and $f_{ij}(x^i)$ depends on the stress tensor on the boundary \cite{deHaro:2000vlm}. 
Similarly, the embedding function for the extremal surface $\mB$ in the bulk is given by 
\begin{equation}\label{eqx1FGd2}
x^i \( z, \sigma^a \) = \ov{0}{x^i}\(\sigma^a \)  + \frac{z^2}{L^2}  \(   \ov{1}{x^i} \( \sigma^a\)  + \ov{1}{y^i} \( \sigma^a\)  \log\( \frac{z}{L}  \) \)+ \mathcal{O}\(\frac{z^4}{L^4}\)\,. 
\end{equation}
From this expansion, we see that the subleading terms are not fully geometric anymore and depend on the details of the boundary state.

Explicitly, performing the CV integral in the vicinity of the brane with $d=2$ yields
\begin{equation}
\begin{split}
\mC^{\rm{sub}}_\mt{V} \(\mA \)&= \max_{\partial \mB={\mA}\cup {\RT}}\[ \frac{  V\( \mathcal{B}_{\mt{L}}\)+V\( \mathcal{B}_{\mt{R}}\) }{\Gbulk \ell}  \] 
\approx \frac{2L V(\wmB)}{\Gbulk \ell} + \mathcal{O}(\zB^0) \,.
\end{split}
\end{equation}
Hence the leading term is still the volume of the island and the subleading contributions are dominated by the upper bound in the radial $z$-integral, \ie these should be included as quantum contributions to the brane complexity. As a result, we will only need to consider the leading contribution, \ie the volume term. 

Now the expression for the effective action given in eq.~\eqref{eq:effectiveaction} does not apply for $d=2$. Rather after a careful examination of the FG expansion and  integration over the radial direction (see section 2.3 in \cite{Chen:2020uac} for more details), the induced action for the $d=2$ brane can be written as
\begin{equation}\label{eq:induct}
I_{\mathrm{induced}} = \frac{1}{16 \pi G_\mt{eff}}\int d^2x\sqrt{-\tilde{g}}
\Big[\frac{2}{\ell_{\rm{eff}}^2} - \tilde{R} \,\log\left(-\frac{L^2
}{2}\tR\right)+\tR +\frac{L^2}{8}\,\tilde{R}^2 +\cdots\Big]\,. \end{equation}
where the two effective scales are
\begin{align}\label{eq:2deff}
\left( \frac{L}{\ell_{\rm{eff}}} \right)^2
=& \ 2\left( 1-4\pi \Gbulk L T_o \right)\;,
&
G_{\rm{eff}}=&\Gbulk/L\,.
\end{align}
The unusual logarithmic term can be understood as arising from the nonlocal Polyakov action induced by the two-dimensional boundary CFT supported by the brane.  

There is a certain degree of ambiguity in how to proceed at this point, but examining our ansatz \reef{eq:defineWgen} for the generalized volume $\widetilde{W}_{\rm gen}(\wmB)$ (with undetermined $\alpha_2$, 
$\beta_2$, $\gamma_2$), we obtain 
\begin{equation}
\mC^{\rm{Island}}_\mt{V,d=2} = \frac{\widetilde{W}_{\rm gen}(\wmB)}{\hat{G}_{\rm eff} \ell} \simeq \frac{1}{\Geff \ell'} \int_{\wmB}  d\sigma\,  \sqrt{{h}} \[ -\frac{\alpha_2}{2}\, \log\left(-\frac{L^2}{2}\tR\right)
+\big(0\big)\,\beta_2+ \gamma_2 \]\,.
\end{equation}  
Here we have ignored any contributions from the $\tilde{R}^2$ and higher terms (denoted by the ellipsis) in eq.~\reef{eq:induct}. We note that these contributions do not contain any UV divergences in the limit $L/\ell_\mt{B}\to0$, and so they can be included as part of the quantum contribution to the complexity. Further, note that tensor contraction multiplying the coefficient $\beta_2$ vanishes for $d=2$. Now the following simple choice of the coefficients,
\beq \label{eq:choice2d}
\alpha_2=0\,,\qquad \gamma_2=2\,, 
\qquad \ell'=\ell\,,
\eeq
yields the desired identification for the two-dimensional complexity
\begin{equation}
\mC^{\rm{sub}}_\mt{V} \(\mA \)\simeq\mC^{\rm{Island}}_\mt{V,d=2} = \frac{\widetilde{W}_{\rm gen}(\wmB)}{\hat{G}_{\rm eff} \ell'}= \frac{ 2\,V(\tilde{\mB})}{\Geff\, \ell'}\,.
\end{equation}

We again note that a similar mismatch from logarithmic divergences in the induced action will appear for any $d=2n$. In this case, no corresponding divergence appears in the holographic complexity in the bulk, which is odd-dimensional. Hence one should only apply eqs.~\reef{eq:ourporposal} and \reef{eq:details} for the generalized CV proposal for higher curvature interactions up to $R^{2n-2}$. A logarithmic divergence will appear at the next order, \ie $R^{2n-1}$, and the corresponding contribution to the complexity will have to be treated separately. Again, it would be interesting to explicitly examine this question in greater detail for higher dimensions.